\documentclass[angeo]{copernicus}

\usepackage{epsfig}
\usepackage[latin2]{inputenc}

\begin{document}

\title{A global study of hot flow anomalies using Cluster multi-spacecraft measurements} 

\author[1]{G.~Facsk\'o\thanks{Now at LPCE, CNRS, Orl\'eans, France}}
\author[1]{Z.~N\'emeth}
\author[1]{G.~Erd\H{o}s}
\author[2]{A.~Kis}
\author[3]{I.~Dandouras}
\affil[1]{KFKI Research Institute for Particle and Nuclear Physics, Budapest, Hungary}
\affil[2]{Geodetic and Geophysical Research Institute, Sopron, Hungary}
\affil[3]{CERS, CNRS, Toulouse, France}
\correspondence{G. Facsk\'o (gfacsko@cnrs-orleans.fr)}

\runningtitle{A global study of HFAs using Cluster multi-spacecraft measurements}
\runningauthor{G.~Facsk\'o et al.}

\maketitle

\begin{abstract}
Hot flow anomalies (HFAs) are studied using observations of the magnetometer and the plasma instrument aboard the four Cluster spacecraft. We study several specific features of tangential discontinuities on the basis of Cluster measurements from the time periods of February-April 2003, December 2005-April 2006 and January-April 2007, when the separation distance of spacecraft was large. The previously discovered condition \citep{facsko08:_statis_study_of_hot_flow} for forming HFAs is confirmed, i.e.~that the solar wind speed and  fast magnetosonic Mach number values are higher than average. Furthermore, this constraint is independent of the \citet{schwartz00:_condit}'s condition for HFA formation. The existence of this new condition is confirmed by simultaneous ACE magnetic field and solar wind plasma observations at the L1 point, at 1.4 million km distance from the Earth. The temperature, particle density and pressure parameters observed at the time of HFA formation are also studied and compared to average values of the solar wind plasma. The size of the region affected by the HFA was estimated by using two different methods. We found that the size is mainly influenced by the magnetic shear and the angle between the discontinuity normal and the Sun-Earth direction. The size grows with the shear and (up to a certain point) with the angle as well. After that point it starts decreasing. The results are compared with the outcome of recent hybrid simulations. 

\end{abstract}

\introduction
\label{sec:intro}

Although \textit{hot flow anomalies} (HFAs), explosive events near the Earth's bow shock have been known more than 20 years \citep{schwartz85, thomsen86:_hot}, their theoretical explanation needs further studies \citep{burgess88:_collid_curren, thomas91:_hybrid, lin02:_global}. The most reliable description of HFAs is so far based on hybrid plasma simulations where electrons are considered as a massless and neutralizing fluid. The original motivation of this work was to verify several predictions presented in \citet{lin02:_global}, but this study led us much further than we expected. In order to do this we determined the size-angle plot (described in the following section). We calculated the related angles and estimated the size in two different ways. \citeauthor{lin02:_global}'s hybrid simulation \citep{lin02:_global} uses a larger simulation box than in other studies mentioned above, and inserts a zero-resistivity surface (magnetopause) to the super-Alfv\'enic plasma flow when the simulation is initialized. This plasma flow moves parallel to the x axis of the box and a shock is formed. A tangential discontinuity is created ahead of the shock, and then the angle between flow direction and normal vector ($\gamma$) can be changed. The simulations were run using different angles and their results suggested that average radius of HFAs is approximately 1-3\,$R_{\mathrm{Earth}}$. A prediction of her theory is that the size of HFAs increases monotonically with $\gamma$ until 80\degree\ and then begins to decrease. Another prediction is that the size of HFAs is a monotonically increasing function of the magnetic field vector direction change angle ($\Delta\Phi$) across the discontinuity \citep{lin02:_global}. The goal of this study was to check the validity of these predictions based on simulation results. 

The four spacecraft Cluster mission provides an excellent opportunity to study HFAs \citep{lucek04:_clust,kecskemety06:_distr_rapid_clust}. We have identified 124 HFAs in the Cluster dataset, which enables a statistical survey. This expands the database of known events since previous analysis was based on significantly fewer events \citep{schwartz00:_condit}. Our results confirm the results of \citet{lin02:_global} that the size depends on the shear and on the angle between the discontinuity normal and Sun-Earth direction as well; furthermore these results strongly support the recently suggested new condition of HFA formation namely that during HFA formation the typical value of the solar wind speed is higher than the average \citep{facsko08:_statis_study_of_hot_flow}. We have used part of \citet{schwartz00:_condit}'s calculations so we have checked his formula (Eq.~\ref{eq:schwartzc}) too. Finally the original purpose led us to confirm the findings of three different previous theories and to discover several new independent condition of HFA formation. 

The structure of this paper is as follows: we first describe the observational methods and the observed events in Section~\ref{sec:data} and \ref{sec:obs}, discuss and present our analysis methods in Section~\ref{sec:discus}, and explain and summarize the result of our study in Section~\ref{sec:conc}.

\section{Data sets}
\label{sec:data}

For our study we used 1\,s and $\left(22.5\,Hz\right)^{-1}$ temporal resolution Cluster FGM (Fluxgate Magnetometer) magnetic field data \citep{balogh01:_clust_magnet_field_inves} and spin averaged time resolution CIS (Cluster Ion Spectrometry) HIA (Hot Ion Analyzer) plasma measurement data \citep{reme01:_first_clust_cis}. We often found the magnetic signatures of the TD -- which interacts with the bow shock and generates the HFA later -- in ACE (Advanced Composition Explorer) MAG (Magnetometer Instrument) 16\,s temporal resolution magnetic field data series \citep{smith98:_ace_magnet_field_exper}. Alfv\'en Mach numbers were calculated and solar wind velocity was determined based on ACE SWEPAM (Solar Wind Electron, Proton, and Alpha Monitor) 16\,s temporal resolution data \citep{mccomas98:_solar_wind_elect_proton_alpha}. ACE SWEPAM data series were used instead of Cluster CIS HIA prime parameter data because in the case of very cold plasmas, as in the solar wind, where thermal velocities are very small compared to the plasma bulk velocity and to the instrument intrinsic energy (and thus velocity) resolution, the relative error in temperature can be large \citep{reme01:_first_clust_cis, caveat_for_data_suppl_by}; furthermore not all the necessary CIS HIA data has been uploaded onto the Cluster Active Archive yet. 

\begin{table*}[t]

\caption{The list of studied  HFA events and spacecraft positions where HFA was observed in GSE system, in $R_{\mathrm{Earth}}$ units. An empty cell indicates that the satellite in question did not observe the magnetic signature of a HFA.} \label{tab:scdataa}
\vskip 4 mm
\centering

\begin{tabular}{|c|c|r@{,}r@{,}r|r@{,}r@{,}r|r@{,}r@{,}r|r@{,}r@{,}r|}
\tophline
date & time & \multicolumn{12}{c|}{s/c positions}  \\
(yymmdd)& (UT) & \multicolumn{3}{c}{C1} & \multicolumn{3}{c}{C2} & \multicolumn{3}{c}{C3} & \multicolumn{3}{c|}{C4} \\ 
\middlehline 
030216 & 10:04 & 10.57 & -1.19 & -9.57 & 11.25 & -0.55 & -9.43 & 11.95 & -0.67 & -9.58 & 12.14 & -0.47 & -9.00 \\
\hline
030216 & 10:48 & 9.82 & -1.45 & -9.66 & 10.53 & -0.78 & -9.54 & 11.27 & -0.90 & -9.69 & 11.46 & -0.73 & -9.12 \\
\hline
030216 & 11:00 & 9.58 & -1.53 & 9.67 & \multicolumn{3}{c|}{} & 11.06 & -0.98 & -9.72 & \multicolumn{3}{c|}{} \\
\hline
030217 & 09:59 & 10.32 & 5.78 & 6.88 & 9.59 & 10.77 & 6.93 & \multicolumn{3}{c|}{} & 10.91 & 5.70 & 6.41 \\
\cline{1-8}\cline{12-14}
030217 & 10:05 & 10.43 & 5.79 & 6.84 & 10.90 & 5.10 & 6.88 & \multicolumn{3}{c|}{} & 11.03 & 5.71 & 6.36 \\
\cline{1-8}\cline{12-14}
030217 & 10:07 & 10.47 & 5.79 & 6.82 & \multicolumn{3}{c}{} & \multicolumn{3}{c}{} & \multicolumn{3}{c|}{} \\
\hline
030221 & 04:18 & 10.43 & -2.08 & -9.60 & 11.17 & -1.49 & -9.46 & 11.85 & -1.67 & -9.62 & 12.06 & -1.50 & -9.03 \\
\hline
030307 & 09:12 & 11.29 & -4.56 & -9.35 & 12.08 & -4.21 & -9.16 & 12.62 & -4.51 & -9.33 & 12.89 & -4.37 & -8.73 \\
\hline
030307 & 09:19 & 11.18 & -4.57 & -9.38 & 11.98 & -4.22 & -9.19 & 12.52 & -4.52 & -9.36 & 12.78 & -4.38 & -8.76 \\
\hline
030307 & 10:15 & \multicolumn{3}{c|}{} & 11.13 & -4.23 & -9.41 & 11.71 & -4.56 & -9.56 & 11.97 & -4.44 & -8.98 \\
\hline
030308 & 12:07 & 12.89 & 1.71 & 6.23 & 13.07 & 0.92 & 6.30 & 12.90 & 1.21 & 5.69 & 13.40 & 1.46 & 5.80 \\
\hline
030317 & 23:57 & 12.51 & -0.22 & 6.42 & 12.55 & -1.03 & 6.48 & 12.41 & -0.70 & 5.88 & 12.95 & -0.55 & 5.97 \\
\hline
030318 & 00:41 & 13.14 & -0.51 & 6.11 & 13.18 & -1.32 & 6.18 & 13.07 & -1.00 & 5.57 & 13.58 & -0.83 & 5.68 \\
\hline
030319 & 06:20 & 10.47 & -6.86 & -9.31 & 11.30 & -6.68 & -9.11 & 11.74 & -7.07 & -9.28 & 12.03 & -6.98 & -8.68 \\
\hline
030319 & 06:52 &  9.96 & -6.78 & -9.44 & \multicolumn{3}{c}{} & \multicolumn{3}{c}{} & \multicolumn{3}{c|}{} \\
\cline{1-5}
030319 & 07:01 & 9.83 & -6.76 & -9.47 & \multicolumn{3}{c}{} & \multicolumn{3}{c}{} & \multicolumn{3}{c|}{} \\
\cline{1-5}
030321 & 15:15 & 10.33 & -7.30 & -9.28 & \multicolumn{3}{c}{} & \multicolumn{3}{c}{} & \multicolumn{3}{c|}{} \\
\hline
030321 & 15:48 &  9.84 & -7.21 & -9.41 & 10.70 & -7.06 & -9.22 & 11.14 & -7.49 & -9.38 & 11.44 & -7.42 & -8.79 \\
\hline
030321 & 16:57 & 8.76 & -6.99 & -9.64 & 9.67 & -6.82 & -9.49 & 10.17 & -7.29 & -9.64 & 10.45 & -7.24 & -9.06 \\
\hline
030321 & 17:12 & 8.52 & -6.93 & -9.68 & 9.44 & -6.76 & -9.54 & 9.96 & -7.25 & -9.68 & 10.22 & -7.19 & -9.10 \\
\hline
030321 & 17:56 & 7.79 & -6.75 & -9.78 & 8.75 & -6.58 & -9.65 & 9.30 & -7.09 & -9.80 & 9.55 & -7.04 & -9.23 \\
\hline
030322 & 19:58 & \multicolumn{3}{c|}{} & 13.84 & -2.92 & 5.67 & 13.79 & -2.62 & 5.05 & 14.28 & -2.47 & 5.19 \\
\hline
030323 & 23:22 &  10.86 & -7.87 & -9.01 & 11.66 & -7.79 & -8.77 & 12.02 & -8.17 & -8.96 & 12.34 & -8.10 & -8.36 \\
\hline
030324 & 00:25 & 9.96 & -7.70 & -9.30 & 10.80 & -7.59 & -9.10 & 11.21 & -8.02 & -9.27 & 11.51 & -7.95 & -8.67 \\
\hline
030324 & 00:57 & 9.50 & -7.59 & -9.43 & 10.36 & -7.48 & -9.24 & 10.79 & -7.93 & -9.40 & 11.08 & -7.87 & -8.81 \\
\hline
030324 & 01:08 & \multicolumn{3}{c}{} & \multicolumn{3}{c|}{} &  10.63 & -7.89 & -9.45 & \multicolumn{3}{c|}{} \\
\cline{1-2}\cline{9-11}
030412 & 01:38 & \multicolumn{3}{c}{} & \multicolumn{3}{c|}{} & 7.76 & -11.04 & -9.44 & \multicolumn{3}{c|}{} \\
\cline{1-2}\cline{9-14}
030412 & 01:42 & \multicolumn{3}{c}{} & \multicolumn{3}{c|}{} & 7.73 & -11.01 & -9.45 & 8.02 & -11.05 & -8.87 \\
\hline
030416 & 16:07 & 8.32 & -12.45 & -8.35 & 8.96 & -12.69 & -8.09 & 9.12 & -13.10 & -8.31 & 9.47& -13.13 & -7.71 \\
\hline
030416 & 16:23 & 8.16 & -12.36 & -8.44 & 8.81 & -12.60 & -8.19 & 8.98 & -13.02 & -8.40 & 9.32 & -13.05 & -7.80 \\
\hline
030416 & 18:18 &  6.90 & -11.58 & -9.09 & 7.65 & -11.82 & -8.87 & 7.85 & -12.33 & -9.04 & 8.17 & -12.37 & -8.45 \\
\hline \hline
051228 & 11:17 &  6.15 &  17.33 &  -3.18 &  7.28 &  17.15 &  -2.32 &  7.39 &  16.48 & -4.19 &   6.97 &  16.49 &  -3.29 \\
\hline
051228 & 12:10 &  6.42 &  17.31 &  -3.69 &  7.52 &  17.13 &  -2.85 & \multicolumn{6}{c|}{} \\
\hline
051228 & 21:51 &  8.41 &  14.50 &  -8.60 &  8.95 &  14.42 &  -7.98 & 9.14 &   13.38 & -9.69 &   9.08 &  14.07 &  -9.09 \\
\hline
051228 & 22:09 &  8.43 &  14.34 &  -8.72 &  8.96 &  14.28 &  -8.10 & 9.14 &   13.22 & -9.81 &   9.10 &  13.93 &  -9.22 \\
\hline
051228 & 22:34 &  8.47 &  14.12 &  -8.89 &  8.96 &  14.05 &  -8.29 & 9.15 &   12.99 & -9.97 &   9.12 &  13.77 &  -9.35 \\
\hline
051228 & 22:39 &  8.47 &  14.08 &  -8.92 &  8.96 &  14.00 &  -8.32 & 9.15 &   12.94 & -10.00&   9.13 &  13.68 &  -9.43 \\
\hline
051229 & 00:01 &  8.55 &  13.30 &  -9.41 &  8.94 &  13.24 &  -8.87 & \multicolumn{6}{c|}{} \\
\cline{1-8}
051229 & 01:20 &  8.57 &  12.47 &  -9.85 & \multicolumn{9}{c|}{} \\
\hline
051229 & 01:54 &  8.56 &  12.09 & -10.01 &  8.82 &  12.04 &  -9.54 & 9.02 &   10.96 & -11.04 &  9.13 &  11.95 & -10.59 \\
\hline
051229 & 02:28 &  8.54 &  11.70 & -10.17 &  8.76 &  11.66 &  -9.71 & \multicolumn{6}{c|}{} \\
\hline \hline
060117 & 04:50 & 11.50 &   7.12 & -10.56 & 11.60 &   7.09 & -10.16 & 11.42 &   5.96 & -11.53 & 11.93 &   6.83 & -11.26 \\
\hline
060126 & 21:22 & 10.25 &   2.38 & -11.00 & 10.09 &   2.49 & -10.76 &  9.77 &   1.44 & -11.83 & 10.69 &   2.28 & -11.84 \\
\hline
060128 & 05:56 & 12.97 &  12.00 &  -1.09 & 13.92 &  11.18 &  -0.17 & 13.78 &  10.71 &  -2.05 & 13.19 &  10.76 &  -1.16 \\
\hline
060128 & 06:12 & 13.09 &  12.00 &  -1.26 & 14.03 &  11.18 &  -0.34 & 13.88 &  10.69 &  -2.22 & 13.32 &  10.75 &  -1.34 \\
\hline
060128 & 07:24 & 13.56 &  11.91 &  -1.98 & \multicolumn{9}{c|}{} \\ 
\hline
060128 & 08:24 & 13.91 &  11.79 &  -2.59 & 14.80 &  11.02 &  -1.68 & 14.62 &  10.42 &  -3.63 & 14.20 &  10.60 &  -2.77 \\
\hline
060128 & 13:23 & 15.00 &  10.67 &  -5.43 & 15.73 &  10.05 &  -4.59 & 15.44 &   9.23 &  -6.56 & 15.33 &   9.61 &  -5.80 \\
\hline
060214 & 22:33 & 10.23 &  -1.38 & -11.02 & 10.18 &  -1.15 & -10.82 &  9.46 &  -2.05 & -11.84 & 10.58 &  -1.59 & -11.88 \\
\hline
\bottomhline
\end{tabular}
\end{table*}

\setcounter{table}{0}

\begin{table*}[t]

\caption{The list of studied  HFA events and spacecraft positions where HFA was observed in GSE system, in $R_{\mathrm{Earth}}$ units. An empty cell indicates that the satellite in question did not observe the magnetic signature of a HFA.} \label{tab:scdatab}
\vskip 4 mm
\centering

\begin{tabular}{|c|c|r@{,}r@{,}r|r@{,}r@{,}r|r@{,}r@{,}r|r@{,}r@{,}r|}
\tophline
date & time & \multicolumn{12}{c|}{s/c positions}  \\
(yymmdd)     & (UT) & \multicolumn{3}{c}{C1} & \multicolumn{3}{c}{C2} & \multicolumn{3}{c}{C3} & \multicolumn{3}{c|}{C4} \\
\middlehline 
060215 & 23:29 & 11.55 &   7.51 &   3.19 & 12.20 &   6.40 &   4.04 & 12.17 &   6.30 &   2.53 & 10.93 &   6.21 &   3.34 \\
\hline
060221 & 01:50 & 17.04 &   5.47 &  -1.77 & 17.53 &   4.36 &  -0.81 & 17.23 &   3.93 &  -2.82 & 16.81 &   4.22 &  -1.96 \\
\hline
060222 & 01:09 & \multicolumn{6}{c|}{}                             &  9.82 &  -3.03 & -11.89 & \multicolumn{3}{c|}{} \\
\hline 
060223 & 04:14 & 13.62 &   5.91 &   2.30 & 14.12 &   4.71 &   3.20 & 14.03 &   4.56 &   1.58 & 13.02 &   4.69 &   2.39 \\
\hline
060310 & 15:31 & 10.84 &  -5.30 & -11.01 & 10.97 &  -5.22 & -10.72 &  9.83 &  -5.76 & -11.91 & 10.84 &  -5.72 & -11.82 \\
\hline
060320 & 04:16 &  9.58 &  -7.01 & -11.07 &  9.75 &  -6.93 & -10.80 &  8.51 &  -7.27 & -11.95 &  9.49 &  -7.41 & -11.89 \\
\hline
060322 & 08:00 & 13.35 &  -7.40 & -10.08 & 13.47 &  -7.71 &  -9.55 & 12.23 &  -8.02 & -11.14 & 12.96 &  -8.01 & -10.80 \\
\hline
060410 & 04:37 & 12.65 & -11.59 &  -8.68 & 12.54 & -12.17 &  -7.96 & 11.34 & -12.11 &  -9.83 & 11.85 & -12.15 &  -9.37 \\
\hline
060410 & 05:27 & 12.23 & -11.59 &  -9.01 & 12.16 & -12.12 &  -8.32 & 10.93 & -12.05 & -10.14 & 11.48 & -12.13 &  -9.71 \\
\hline
060410 & 07:53 & 10.89 & -11.46 &  -9.85 & 10.90 & -11.85 &  -9.27 &  9.61 & -11.75 & -10.94 & 10.23 & -11.92 & -10.59 \\
\hline
060410 & 08:28 & 10.54 & -11.40 & -10.03 & 10.57 & -11.75 &  -9.47 &  9.26 & -11.64 & -11.10 &  9.91 & -11.84 & -10.78 \\
\hline
060410 & 12:52 &  7.54 & -10.49 & -10.98 &  7.72 & -10.59 & -10.64 & \multicolumn{6}{c|}{} \\
\cline{1-8}
060416 & 12:39 & 13.07 &  -7.25 &   1.87 & 12.31 &  -8.49 &   1.06 & \multicolumn{6}{c|}{} \\
\hline
060416 & 12:45 & 13.11 &  -7.33 &   1.81 & 12.40 &  -8.37 &   2.81 & 12.34 &  -8.57 &   0.99 & 11.92 &  -7.82 &   1.71 \\
\hline
060416 & 13:22 & 13.34 &  -7.72 &   1.44 & 14.04 &  -9.21 &  -0.12 & 13.07 & -10.38 &  -1.08 & 12.88 &  -9.82 &  -0.36 \\
\hline
060416 & 16:30 & 14.14 &  -9.51 &  -0.48 & 13.46 & -10.54 &   0.57 & 13.14 & -10.68 &  -1.47 & 12.99 & -10.13 & -0.74 \\
\hline
060416 & 16:39 & 14.16 &  -9.58 &  -0.57 & 13.48 & -10.59 &   0.50 & 13.16 & -10.73 &  -1.54 & 13.01 & -10.19 & -0.82 \\
\hline
060416 & 18:32 & 14.34 & -10.46 &  -1.71 & 13.70 & -11.46 &  -0.66 & 13.25 & -11.56 &  -2.77 & 13.21 & -11.11 & -2.06 \\
\hline
060416 & 20:01 & 14.35 & -11.06 &  -2.60 & 13.75 & -12.04 &  -1.56 & 13.19 & -12.10 &  -3.70 & 13.23 & -11.72 & -3.01 \\
\hline \hline
070104 & 03:53 & 10.15 &  12.49 & -10.41 & 10.59 &  12.74 &  -9.78 & 10.69 &  11.67 & -11.24 &  \multicolumn{3}{c|}{ } \\
\hline
070104 & 04:36 & 10.13 &  12.06 & -10.57 & 10.53 &  12.34 &  -9.99 & 10.64 &  11.25 & -11.40 &  10.64 & 11.31 & -11.38 \\
\hline
070104 & 06:20 & 10.01 &  10.96 & -10.91 & 10.34 &  11.30 & -10.44 & 10.44 &  10.20 & -11.72 &  10.45 & 10.26 & -11.71 \\
\hline
070104 & 05:08 & 10.10 &  11.74 & -10.69 & 10.48 &  12.03 & -10.14 & 10.59 &  10.95 & -11.51 &  10.59 & 11.00 & -11.49 \\
\hline
070106 & 16:07 & 10.39 &  10.06 & -11.01 & 10.69 &  10.41 & -10.60 & \multicolumn{6}{c|}{ } \\
\hline
070108 & 11:25 & 10.17 &  15.81 &  -6.43 & 11.03 &  15.52 &  -5.28 & 11.08 &  14.82 &  -7.27 &  11.06 & 14.84 & -7.22 \\ 
\hline
070116 & 09:40 & 10.08 &   4.54 & -11.18 & 10.19 &   5.02 & -11.14 & 10.11 &   3.97 & -11.93 &  10.15 &  4.03 & -11.93 \\
\hline
070116 & 10:00 &  9.91 &   4.27 & -11.15 & 10.00 &   4.77 & -11.13 &  9.93 &   3.73 & -11.89 &   9.97 &  3.79 & -11.89 \\
\hline
070116 & 10:49 &  9.48 &   3.63 & -11.04 &  9.54 &   4.15 & -11.08 &  9.48 &   3.14 & -11.77 &   9.53 &  3.21 & -11.78 \\
\hline
070117 & 16:38 & 10.35 &  14.55 &  -2.64 & 11.24 &  13.79 &  -1.34 & 11.28 &  13.50 &  -3.37 &  11.24 & 13.50 & -3.317 \\
\hline
070118 & 07:49 & 13.27 &  10.98 &  -9.65 & 13.84 &  10.90 &  -8.91 & 13.69 &   9.91 & -10.53 &  13.69 &  9.95 & -10.50 \\
\hline
070118 & 09:42 & 13.09 &  10.01 & -10.20 & 13.59 &  10.03 &  -9.56 & 13.43 &   9.00 & -11.06 &  13.44 &  9.05 & -11.04 \\
\hline
070118 & 12:13 & 12.65 &   8.58 & -10.77 & 13.05 & 8.73 & -10.28 & 12.89 & 7.66 & -11.61 & 12.91 & 7.71 & -11.59 \\
\hline
070118 & 14:35 & 12.01 &   7.08 & -11.12 & 12.31 & 7.34 & -10.79 & 12.15 & 6.26 & -11.93 & 12.18 & 6.32 & -11.91 \\
\hline
070118 & 19:34 &  9.83 &   3.48 & -11.08 &  9.91 & 3.98 & -11.10 &  9.83 & 2.98 & -11.82 &  9.87 & 3.04 & -11.82 \\
\hline
070120 & 18:18 & 13.58 &   9.71 & -10.06 & 13.90 & 8.66 & -10.94 & 13.92 & 8.71 & -10.90 & \multicolumn{3}{c|}{} \\
\cline{1-11}
070130 & 16:44 & 10.72 &   1.63 & -11.13 & 10.85 & 2.01 & -11.13 & \multicolumn{6}{c|}{}  \\
\hline
070201 & 06:48 &  15.78 & 10.20 &  -6.63 & 16.48 & 9.54 &  -5.61 & 16.23 & 8.88 &  -7.56 & 16.22 & 8.91 &  -7.51 \\
\hline
070201 & 22:07 &  13.05 &  3.36 & -11.2  & 13.32 & 3.52 & -10.95 & 12.92 & 2.53 & -12.00 & 12.97 & 2.59 & -11.99 \\
\hline
070201 & 22:16 &  12.97 &  3.27 & -11.21 & 13.23 & 3.44 & -10.97 & 12.83 & 2.45 & -12.01 & 12.88 & 2.51 & -12.00 \\
\hline
070202 & 01:31 &  11.02 &  1.36 & -11.17 & 11.17 & 1.71 & -11.14 & 10.82 & 0.75 & -11.90 & 10.88 & 0.80 & -11.90 \\
\hline
070209 & 02:14 & \multicolumn{6}{c|}{}   & 12.68 & 0.53 & -12.04 & \multicolumn{3}{c|}{} \\
\hline
070215 & 01:35 &  13.61 &  8.15 &  -0.21 & 14.10 & 6.94 &   0.93 & 14.16 & 6.88 &  -0.96 & 14.10 & 6.89 &  -0.89 \\
\hline
070215 & 02:29 &  14.21 &  8.13 &  -0.74 & \multicolumn{3}{c|}{} & 14.74 & 6.82 &  -1.53 & 14.68 & 6.83 &  -1.46 \\
\cline{1-5}\cline{9-14}
070215 & 02:49 &  14.41 &  8.11 &  -0.94 & \multicolumn{3}{c|}{} & 14.93 & 6.79 &  -1.73 & 14.88 & 6.80 &  -1.67 \\
\cline{1-5}\cline{9-14}
070215 & 03:13 & 14.65 & 8.08 &  -1.17 & \multicolumn{9}{c|}{} \\
\hline
070215 & 03:55 & 15.05 & 8.02 & -1.58 & 15.56 & 6.84 & -0.43 & 15.53 & 6.67 & -2.41 & 15.49 & 6.68 & -2.35 \\
\hline
070215 & 04:01 & 15.10 & 8.01 & -1.64 & 15.62 & 6.83 & -0.48 & 15.58 & 6.65 & -2.47 & 15.54 & 6.67 & -2.41 \\
\hline
070215 & 08:44 & 17.02 & 7.22 & -4.29 & 17.57 & 6.18 & -3.20 & 17.33 & 5.79 & -5.23 & 17.31 & 5.81 & -5.17 \\
\hline
070215 & 15:16 & 17.86 & 5.35 & -7.49 & 18.36 & 4.60 & -6.59 & 17.89 & 3.97 & -8.47 & 17.90 & 4.01 & -8.42 \\
\hline
\bottomhline
\end{tabular}
\end{table*}

\setcounter{table}{0}

\begin{table*}[t]

\caption{The list of studied  HFA events and spacecraft positions where HFA was observed in GSE system, in $R_{\mathrm{Earth}}$ units. An empty cell indicates that the satellite in question did not observe the magnetic signature of a HFA.} \label{tab:scdatac}
\vskip 4 mm
\centering

\begin{tabular}{|c|c|r@{,}r@{,}r|r@{,}r@{,}r|r@{,}r@{,}r|r@{,}r@{,}r|}
\tophline
date & time & \multicolumn{12}{c|}{s/c positions}  \\
(yymmdd)     & (UT) & \multicolumn{3}{c}{C1} & \multicolumn{3}{c}{C2} & \multicolumn{3}{c}{C3} & \multicolumn{3}{c|}{C4} \\
\middlehline 
070301 & 04:56 & 12.47 & 5.00 &  1.65 & 12.64 & 3.71 &  2.67 & 12.88 & 3.76 &  0.98 & 12.80 & 3.77 &  1.04 \\
\hline
070301 & 07:10 & 14.38 &  4.72 &   0.34 & 14.61 & 3.40 &   1.40 & 14.72 & 3.37 & -0.44 & 14.65 & 3.39 & -0.37 \\
\hline
070301 & 09:43 & 16.06 &  4.24 &  -1.16 & 16.32 & 2.93 &   -0.08 & 16.29 & 2.81 & -2.03 & 16.24 & 2.83 & -1.96 \\
\hline
070301 & 10:30 & 16.49 &  4.07 &  -1.63 & \multicolumn{9}{c|}{} \\
\hline
070302 & 02:03 & 17.87 & -0.58 &  -9.15 & 18.12 & -1.23 &  -8.49 & 17.40 & -1.80 & -10.15 & 17.44 & -1.77 & -10.10 \\
\hline
070313 & 05:36 & 15.61 &  1.36 &  -0.24 & 15.65 & -0.02 &   0.77 & 15.68 & -0.09 &  -1.11 & 15.63 & -0.05 & -1.04 \\
\hline
070314 & 07:53 & 13.15 & -5.91 & -11.14 & 13.21 & -6.06 & -10.90 & 12.31 & -6.54 & -11.98 & 12.38 & -6.53 & -11.97 \\
\hline
070314 & 08:36 & 12.64 & -6.00 & -11.21 & 12.70 & -6.10 & -11.00 & 11.81 & -6.58 & -12.03 & 11.88 & -6.57 & -12.02 \\
\hline
070314 & 12:51 &  9.18 & -6.27 & -11.17 &  9.21 & -6.06 & -11.17 &  8.40 & -6.56 & -11.85 &  8.49 & -6.56 & -11.86 \\
\hline
070314 & 15:52 &  6.20 & -6.10 & -10.53 &  6.20 & -5.68 & -10.67 &  5.52 & -6.18 & -11.09 &  5.63 & -6.19 & -11.12 \\
\hline
070315 & 12:14 & 13.84 &   1.56 & 1.21 & \multicolumn{3}{c|}{} & 14.01 & 0.18 & 0.43 & 13.93 & 0.22 & 0.50 \\
\cline{1-5}\cline{9-14}
070316 & 18:13 & 12.02 &  -6.57 & -11.25 & \multicolumn{3}{c|}{} & 11.16 & -7.09 & -12.06 \\
\hline
070316 & 19:56 & 10.69 &  -6.67 & -11.30 & 10.72 &  -6.61 & -11.20 &  9.85 &  -7.07 & -12.05 &  9.93 & -7.07 & -12.04 \\
\hline
070319 & 03:39 & 11.55 &  -7.10 & -11.28 & 11.57 &  -7.14 & -11.12 & 10.64 &  -7.55 & -12.08 & 10.72 & -7.55 & -12.07 \\
\hline
070319 & 04:27 & 10.93 &  -7.12 & -11.31 & 10.95 &  -7.10 & -11.19 & \multicolumn{6}{c|}{} \\
\hline
070328 & 13:41 & 12.01 &  -9.01 & -11.06 & 11.94 &  -9.23 & -10.79 & 10.97 &  -9.47 & -11.93 & 11.05 & -9.48 & -11.91 \\
\hline
070328 & 15:22 & 10.84 &  -8.95 & -11.24 & 10.78 &  -9.04 & -11.06 &  9.82 &  -9.30 & -12.06 &  9.90 & -9.31 & -12.05 \\
\hline
070328 & 16:07 & 10.29 &  -8.89 & -11.29 & 10.23 &  -8.92 & -11.14 &  9.28 &  -9.19 & -12.08 &  9.36 & -9.20 & -12.07 \\
\hline
070328 & 16:50 &  9.74 &  -8.82 & -11.31 &  9.70 &  -8.80 & -11.19 &  8.75 &  -9.07 & -12.08 &  8.84 & -9.09 & -12.07 \\
\hline
070429 & 20:40 & 11.79 &  -11.03 & -1.00 & 10.88 & -12.27 &  -0.21 & 10.79 & -12.27 &  -2.12 & 10.79 & -12.21 & -2.05 \\
\hline
070429 & 21:00 & 11.84 &  -11.24 & -1.20 & 10.92 & -12.47 &  -0.40 & 10.82 & -12.47 &  -2.32 & 10.82 & -12.40 & -2.24 \\
\hline
070429 & 22:05 & 11.97 &  -11.89 & -1.83 & 11.04 & -13.01 &  -1.04 & 10.88 & -13.07 &  -2.99 & 10.89 & -13.01 & -2.91 \\
\hline
070429 & 23:02 & 12.04 &  -12.41 & -2.37 & 11.10 & -13.58 &  -1.57 & 10.89 & -13.53 &  -3.54 & 10.89 & -13.47 & -3.46 \\
\hline
070430 & 02:01 & 11.98 &  -13.79 & -4.03 & \multicolumn{9}{c|}{} \\
\hline
\bottomhline
\end{tabular}
\end{table*}

\begin{figure}[!th]
\centering
\epsfig{file=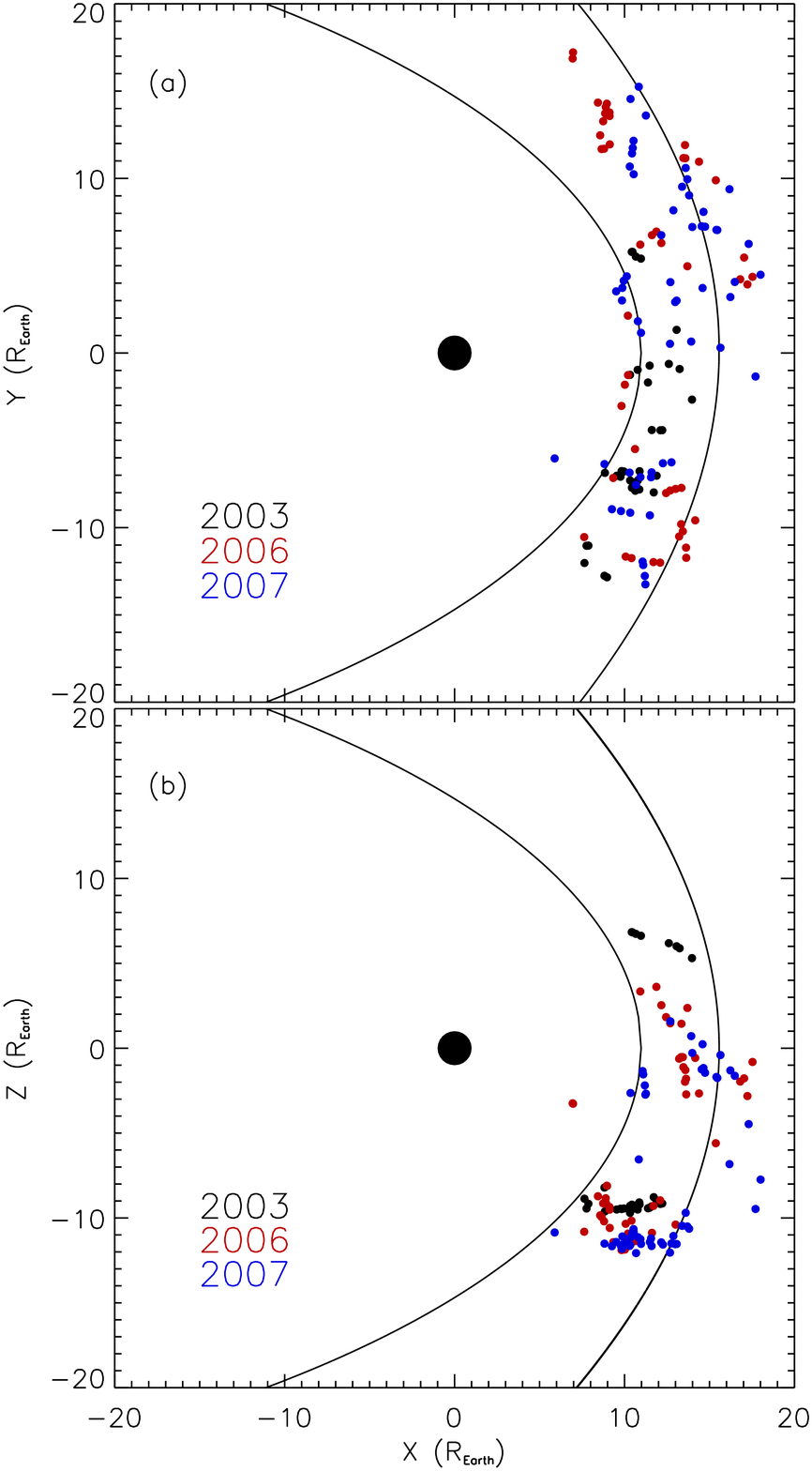,width=200pt}  
\caption{HFA locations (a)  in XY GSE and (b) XZ GSE plane projections and the average bow shock and magnetopause positions. The coordinates were plotted in units of $R_{\mathrm{Earth}}$. The shapes of the magnetopause and the bow shock were calculated with the average solar wind pressure \citep{sibeck91:_solar, tsyganenko95:_model} and Alfv\'en-Mach number during HFA formation \citep{peredo95:_three_alfven_mach}. The black, red and blue points show Cluster positions when HFAs were observed in 2003, 2006 and 2007, respectively.}
\label{fig:positions}
\end{figure}

\begin{figure}[!ht]
\centering
\epsfig{file=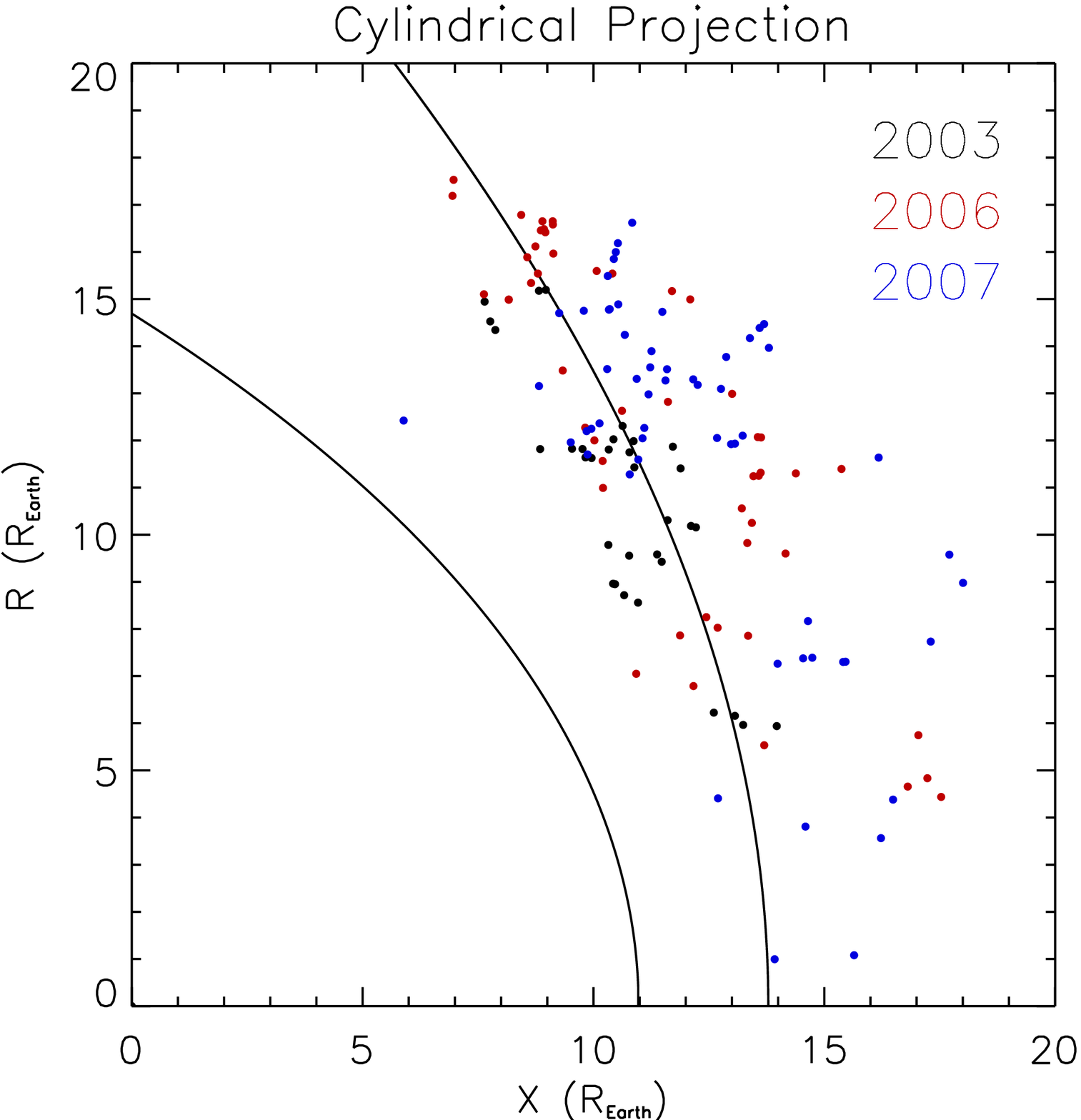,width=200pt}  
\caption{Cylindrical projection of Cluster SC center positions during HFA observation and the average bow shock and magnetopause positions in GSE system. The shape of the magnetopause and the bow shock were calculated using the average solar wind pressure \citep{sibeck91:_solar, tsyganenko95:_model, peredo95:_three_alfven_mach}. The black, red and blue points show the Cluster SC positions when HFA was observed in 2003, 2006 and 2007, respectively. The coordinates were plotted in $R_{\mathrm{Earth}}$ units.}
\label{fig:cyl}
\end{figure}

We set a series of criteria for the selection of HFA events based on \citet{thomsen86:_hot, thomsen93:_obser_test_hot_flow_abnom, sibeck99:_compr, sibeck02:_wind} that were:
\begin{enumerate}
\item The rim of the cavity must be visible as a sudden increase of magnetic field magnitude compared to the unperturbed solar wind region's value. Inside the cavity the magnetic field strength drops and its direction turns around. 
\item The solar wind speed drops and its direction always turns away from the Sun-Earth direction. 
\item The solar wind temperature increases and its value reaches up to several ten million Kelvin degrees.
\item The solar wind particle density also increases on the rim of the cavity and drops inside the HFA.
\end{enumerate}
Using these criteria we identified 124 events in the 2003, 2006 and 2007 data. Two of these events were studied by \citet{kecskemety06:_distr_rapid_clust}, and a statistical study of 33 events in the 2003 data was analyzed by \citet{facsko08:_statis_study_of_hot_flow}. The positions of the events are given in Tab.~\ref{tab:scdataa} and Fig.~\ref{fig:positions}. All of them were observed beyond the bow shock in the February-April, 2003, December 2005-April 2006 and January-April, 2007 time intervals. A fraction of these events was located very far from the bow shock and the Earth ($\ge 19\,R_{\mathrm{Earth}}$), occurring mainly in 2007. Only the position of tetrahedron center of the Cluster SC is plotted in Fig.~\ref{fig:positions},~\ref{fig:cyl} because the length of the orbital section is comparable with the thickness of the lines drawn. The bow shock position was calculated using the average Alfv\'en Mach number during formation of the events ($M_A=11.8$, Sec.~\ref{sec:speeddistr}) according to the model described in \citet{peredo95:_three_alfven_mach}. The position of the magnetopause was calculated using the same average solar wind pressure ($1.73\pm0.8$\,nPa, Sec.~\ref{sec:pressure}) as that in \citet{sibeck91:_solar} and \citet{tsyganenko95:_model}. 

The cylindrical projection of the center of the Cluster SC positions is also plotted to more easily determine whether the observations were performed beyond or inside the average bow shock (Fig.~\ref{fig:cyl}). Fig.~\ref{fig:positions} seems to indicate that the HFAs are mostly located within the magnetosheath, with some inside the magnetosphere. However, this is only a feature of the applied projection. The position of the bow shock was calculated using the average solar wind pressure during, the HFA event. All HFAs were beyond the actual bow shock when we observed them. However the bow shock position changes quickly, presenting explanation for why some of the events seem to be located in the magnetosheath. 

\section{Analysis}
\label{sec:obs}

\subsection{Size-angle plots}
\label{sec:sizeangleplots}

The main purpose of this paper is to determine experimentally the role the different angles ($\gamma$, $\Delta\Phi$) play in controlling HFA size. In the next two sections we therefore calculate the angles associated with each HFA and its size. 

\subsubsection{Determination of angles}
\label{sec:angles}

\begin{table*}[t]

\caption{Parameters of TD normal vectors: $\lambda_{\mathrm{2}}/\lambda_{\mathrm{3}}$ is the ratio of 2nd and 3rd eigenvalues, $B_{\mathrm{min}}$ is the smallest magnetic field component in minimum variance system, $\Delta\mathbf{n}$ is the error cone of minimum variance method, $\gamma$ is the angle between the Sun direction and TD normal, $\Delta\Phi$ is the direction change across the discontinuity and $\theta$ the angle between the bow shock normal and the $\mathbf{B}$ magnetic field vector. Boldface letter shows quasi-perpendicular conditions; the angles were calculated by scaling a model BS to the location of Cluster-1 and 3 spacecraft.} \label{tab:tddataa}
\vskip 4 mm
\centering

\begin{tabular}{|c|c|c|r@{,}r@{,}r|r@{,}r@{,}r|c|c|c|c|c|r@{,}r|}
\tophline
date & time & s/c & \multicolumn{3}{c|}{$\mathbf{n}_{\mathbf{B}_{\mathrm{u}} \times \mathbf{B}_{\mathrm{d}}}$} & \multicolumn{3}{c|}{$\mathbf{n}_{\mathrm{minvar}}$} & $\frac{\lambda_{\mathrm{2}}}{\lambda_{\mathrm{3}}}$ & $B_{\mathrm{min}}$ & $\Delta\mathbf{n}$ & $\gamma$ & $\Delta\Phi$ &  \multicolumn{2}{c|}{$\theta_{\mathrm{C1, C3}}$} \\
(yymmdd) & (UT) & & \multicolumn{3}{c|}{} & \multicolumn{3}{c|}{} & & $\left(nT\right)$ & $\left({}^{o}\right)$ & $\left({}^{o}\right)$ & $\left({}^{o}\right)$ & \multicolumn{2}{c|}{$\left({}^{o}\right)$} \\
\middlehline 
030216 & 10:04 & C1 &  0.53&-0.70& 0.47 & 0.40&-0.69&0.61 & 1.1 & 1.50 & 76.53 & 66 & 51 &  27 &  27 \\
\hline
030216 & 10:48 & C1 & -0.06& 0.39& 0.92 & -0.06&0.39&0.92 & 4.0 & 0.00 & 8.18 & 93 & 73 & 37 &  45  \\
         &       & C2 & \multicolumn{3}{c|}{} & -0.29& 0.37& 0.88 & 1.9 & 1.50 & 13.74 & & & \multicolumn{2}{c|}{ } \\
         &       & C3 & \multicolumn{3}{c|}{} &  0.12& 0.44& 0.89  & 1.6 & -0.63 & 18.65 & & & \multicolumn{2}{c|}{ } \\
         &       & C4 & \multicolumn{3}{c|}{} & -0.11& 0.41& 0.91 & 2.0 & 0.07 & 13.84 & & & \multicolumn{2}{c|}{ } \\
\hline
030216 & 11:00 & ACE & -0.21&-0.98&-0.03 & 0.10&0.98&0.20 & 2.0 & 0.86 & 30.95 & 98 & 42 & \multicolumn{2}{c|}{ } \\
\hline
030216 & 11:02 & ACE & 0.19&-0.09&0.98 & 0.21&-0.08&0.97 & 1.7 & -0.11 & 32.67 & 80 & 61 & \textbf{48} &  \textbf{48} \\
\hline
030217 & 09:59 & ACE & -0.46&0.18&0.87 & -0.63&0.23&0.74 & 1.7 & 0.82 & 31.71 & 99 & 19 & 14 &  13 \\
\hline
030217 & 10:05 & ACE & 0.70&0.63&-0.33 & 0.70&0.63&-0.33 & 8.5 & -0.02 & 10.05 & 48 & 73 & 31 & 28 \\
\hline
030217 & 10:07 & ACE & 0.66&0.75&0.05 & 0.67&0.74&0.05 & 4.7 & -0.08 & 23.32 & 54 & 63 & 32 &  29 \\
\hline
030217 & 10:08 & ACE & 0.17&0.48&-0.86 & -0.06&-0.68&0.73 & 9.8 & 1.24 & 41.98 & 82 & 53 &  31 & 28 \\
\hline
030221 & 04:18 & C1 &  \multicolumn{3}{c|}{} & 0.71&-0.66&-0.25 & 5.2 & -1.24 & 7.91 && &  17 & 21 \\
         &       & C2 & \multicolumn{3}{c|}{} & 0.71&-0.66&-0.24 & 3.8 & -1.18 & 9.24 && & \multicolumn{2}{c|}{ } \\
         &       & C3 & \multicolumn{3}{c|}{} & 0.76&-0.62&-0.19 & 4.2 & -0.80 & 8.76 && & \multicolumn{2}{c|}{ } \\
         &       & C4 & -0.67&0.73&-0.12 & 0.73&-0.62&-0.27 & 5.8 & -1.08 & 7.33 & 96 & 9 & \multicolumn{2}{c|}{ } \\
\hline
030307 & 09:12 & ACE & 0.81&0.22&-0.54 & 0.80&0.35&-0.50 & 1.2 & -0.76 & 75.06 & 66 & 30 & \textbf{77} & \textbf{72} \\
\hline
030307 & 09:19 & ACE & 0.72&0.41&-0.55 & 0.95&0.06&-0.31 & 1.2 & 0.97 & 59.14 & 85 & 7 &  \textbf{63} &  \textbf{71} \\
\hline
030307 & 10:15 & ACE & \multicolumn{3}{c|}{} & 0.61&0.39&0.69 & 1.1 & 0.12 & 78.12 & \multicolumn{2}{c|}{}  &  \textbf{67} &  \textbf{68} \\*
         &       & C2 & -0.53&-0.43&-0.73 & 0.22&0.75&-0.62 & 1.8 & 0.61 & 15.55 & \multicolumn{2}{c|}{} & \multicolumn{2}{c|}{ } \\*
         &       & C3 & \multicolumn{3}{c|}{} & 0.43&0.50&0.76 & 1.8 & 0.01 & 19.85 & \multicolumn{2}{c|}{} & \multicolumn{2}{c|}{ } \\*
         &       & C4 & \multicolumn{3}{c|}{} & 0.17&0.79&-0.60 & 1.5 & 0.12 & 26.17 & \multicolumn{2}{c|}{} & \multicolumn{2}{c|}{ } \\
\hline
030308 & 12:07 & ACE & \multicolumn{3}{c|}{} & 0.56&0.38&0.73 & 1.7 & 0.00 & 34.32 & & &  \textbf{66} & \textbf{58} \\
         &       & C4 & -0.36&-0.35&-0.87 & 0.54&0.30&0.78 & 1.8 & 0.68 & 17.86 & 111 & 87 &  30 & 27 \\
\hline
030317 & 23:57 & C4 & 0.81&0.33&-0.48 & 0.89&0.25&-0.38 & 4.3 & -1.13 & 10.93 & 61 & 37 & \multicolumn{2}{c|}{ } \\
\hline
030318 & 00:41 & ACE & 0.62&0.75&0.23 & 0.51&0.80&0.32 & 2.3 & 1.09 & 25.83 & 67 & 40 &  26 & 29 \\
\hline
030319 & 06:20 & ACE & 0.27&-0.73&0.63 & 0.18&-0.71&0.67 & 1.4 & 0.38 & 44.64 & 79 & 44 &  8 & 16 \\
\hline
030319 & 06:52 & ACE & -0.29&-0.37&-0.88 & 0.38&0.30&0.87 & 1.3 & -0.24 & 53.73 & 95 & 19 & 34 & \textbf{47} \\
\hline
030319 & 07:01 & ACE & -0.67&0.31&-0.68 & -0.71&0.58&-0.40 & 5.9 & 0.06 & 12.77 & 93 & 4 &  34 & \textbf{47} \\
\hline
030321 & 15:15 & ACE & -0.60&0.10&-0.79 & 0.60&-0.19&0.78 & 1.9 & -0.13 & 27.8 & 119 & 54 & \multicolumn{2}{r|}{27} \\
\hline
030321 & 15:48 & ACE & \multicolumn{3}{c|}{} & 0.71& 0.07&0.70 & 1.7 & -0.21 & 13.75  & & & 26 & 27 \\
         &       & C4 & 0.78&0.27&0.57 & 0.78&0.27&0.57 & 3.1 & 0.00 & 23.62 & 51 & 54 & \multicolumn{2}{c|}{ } \\
\hline
030321 & 16:57 & ACE & 0.43&0.73&0.53 & 0.40&0.76&0.52 & 2.5 & 0.12 & 24.91 & 73 & 42 & 24 & 22 \\
\hline
030321 & 17:12 & ACE & \multicolumn{3}{c|}{} & 0.55&-0.34&0.76 & 6.0 & 0.08 & 13.22 & & & 39 &35  \\
         &       & C3 & 0.60&-0.29&0.75 & 0.64&-0.29&0.71 & 11.4 & -0.24 & 6.27 & 53 & 92 & \multicolumn{2}{c|}{ } \\
         &       & C4 & \multicolumn{3}{c|}{} & 0.58&-0.32&0.74 & 3.9 & -0.40 & 12.08  & & & \multicolumn{2}{c|}{ } \\
\hline
030321 & 17:56 & ACE & -0.13&0.19&0.97 & 0.77&0.23&0.59 & 4.2 & -0.41 & 16.10 & 95 & 47 & \textbf{81} & \textbf{84} \\
\hline
030322 & 19:58 & C4 & 0.43&-0.15&-0.89 & -0.55&0.25&0.80 & 1.0 & 1.16 & 87.27 & 78 & 30 & 29 & 32 \\
\hline
030323 & 23:22 & ACE & \multicolumn{3}{c|}{} & 0.14&0.86&0.49 & 2.4 & -0.09 & 19.81 & & & 19 & 26 \\
         &       & C3 & 0.46&0.87&0.17 & 0.36 &0.90 & 0.23 & 3.2 & 0.44 & 10.26 & 63 & 80 & \multicolumn{2}{c|}{ } \\
         &       & C4 & \multicolumn{3}{c|}{} & 0.32&0.91&0.24 & 1.9 & 0.36 & 16.23 & & & \multicolumn{2}{c|}{ } \\
\hline
030324 & 00:25 & ACE & 0.82&-0.42&0.40 & 0.93&-0.34&0.12 & 14.1 & -0.84 & 8.89 & 82 & 10 & 36 & 37 \\
\hline
030324 & 00:57 & C2 & -0.83&-0.46&0.30 & 0.83&0.47&-0.30 & 1.2 & 0.05 & 35.48 & \multicolumn{2}{c|}{} & 16 & 17 \\
\hline
030324 & 01:08 & ACE & -0.06&0.44&-0.90 & -0.10&-0.25&0.96 & 3.8 & 0.43 & 18.05 & 93 & 107 & 19 & 16 \\
\hline
030412 & 01:38 & ACE & -0.48&-0.29&-0.83 & 0.67&0.15&0.72 & 9.0 & -0.95 & 8.72 & 119 & 88 & 33 & 34 \\
\hline
030412 & 01:42 & ACE & 0.48&0.28&0.83 & 0.56&0.20&0.80 & 3.8 & -0.52 & 17.02 & 76 & 31 & 35 & \textbf{46} \\
\hline
030416 & 16:07 & ACE & -0.44&-0.52&0.73 & -0.05&-0.75&0.66 & 6.9 & -1.28 & 11.34 & 112 & 123 & 18 & 16 \\
\hline
030416 & 16:23 & ACE & 0.23&0.18&-0.96 & -0.25&-0.19&0.95 & 1.7 & 0.11 & 30.52 & 83 & 30 & 18 & 16 \\
\hline
030416 & 18:18 & ACE & 0.56&0.82&-0.12 & 0.75&0.59&-0.29 & 7.6 & -1.43 & 10.42 & 57 & 101 & 17 & 15 \\
\hline \hline
051228 & 11:18   & C1  & -0.82 & -0.39 &  0.42 & -0.87 &  0.41 & -0.29 &  1.2 &  0.09 & 39.66 & 144 &   127 & \textbf{68} & \textbf{73} \\
\hline
\bottomhline
\end{tabular}
\end{table*}

\setcounter{table}{1}

\begin{table*}[t]

\caption{Parameters of TD normal vectors: $\lambda_{\mathrm{2}}/\lambda_{\mathrm{3}}$ is the ratio of 2nd and 3rd eigenvalues, $B_{\mathrm{min}}$ is the smallest magnetic field component in minimum variance system, $\Delta\mathbf{n}$ is the error cone of minimum variance method, $\gamma$ is the angle between the Sun direction and TD normal, $\Delta\Phi$ is the direction change across the discontinuity and $\theta$ the angle between the bow shock normal and the $\mathbf{B}$ magnetic field vector. Boldface letter shows quasi-perpendicular conditions; the angles were calculated by scaling a model BS to the location of Cluster-1 and 3 spacecraft.} \label{tab:tddatab}
\vskip 4 mm
\centering

\begin{tabular}{|c|c|c|r@{,}r@{,}r|r@{,}r@{,}r|c|c|c|c|c|r@{,}r|}
\tophline
date & time & s/c & \multicolumn{3}{c|}{$\mathbf{n}_{\mathbf{B}_{\mathrm{u}} \times \mathbf{B}_{\mathrm{d}}}$} & \multicolumn{3}{c|}{$\mathbf{n}_{\mathrm{minvar}}$} & $\frac{\lambda_{\mathrm{2}}}{\lambda_{\mathrm{3}}}$ & $B_{\mathrm{min}}$ & $\Delta\mathbf{n}$ & $\gamma$ & $\Delta\Phi$ &  \multicolumn{2}{c|}{$\theta_{\mathrm{C1, C3}}$} \\
(yymmdd) & (UT) & & \multicolumn{3}{c|}{} & \multicolumn{3}{c|}{} & & $\left(nT\right)$ & $\left({}^{o}\right)$ & $\left({}^{o}\right)$ & $\left({}^{o}\right)$ & \multicolumn{2}{c|}{$\left({}^{o}\right)$} \\
\middlehline 
051228 & 21:50  &  ACE & -0.73 &  0.16 & -0.66 &  0.85 & -0.18 &  0.50 &  2.5 &  0.64 & 19.57 &   137 &     7 & \textbf{60} & \textbf{59} \\
\hline
051228 & 22:10  &  C1 & \multicolumn{3}{c|}{ } &  0.05 &  0.76 & -0.65  &  2.6  &   0.33 &     17.07  & & &  43 & 40 \\
       & 22:10  &  C2 & \multicolumn{3}{c|}{ }   &   0.18 &  0.79 & -0.59  &  1.5  &   0.92  &  30.05  & & & \multicolumn{2}{c|}{} \\
       & 22:10  &  C3 &  0.03 &  0.87 & -0.49 &  0.04 &  0.84 & -0.54 &  4.8 &  0.19 & 10.86 &    88 &    15 & \multicolumn{2}{c|}{ } \\
\hline
051228 & 22:20  &  C1 & \multicolumn{3}{c|}{ } &  0.23  & 0.82 &  -0.52 &   1.9  &  -0.49 &   16.48  & & &  $\textbf{63}$ &  \textbf{53} \\
       & 22:20  &  C2 & -0.25 & -0.85 &  0.46 &  0.13 &  0.85 & -0.51 &  2.2 & -0.46 & 14.28 &   104 &    97 & \multicolumn{2}{c|}{ } \\
       & 22:20 &   C3 & \multicolumn{3}{c|}{ }  &   0.29 &  0.84 & -0.46 &   2.2 &    0.16 &     14.65  & & & \multicolumn{2}{c|}{} \\
       & 22:20  &  C4 & \multicolumn{3}{c|}{ } &  0.34  & 0.81 & -0.48 &   1.4 &    0.11 &     27.67  & & & \multicolumn{2}{c|}{} \\
\hline
051228 & 22:35  &  ACE & -0.64 & -0.75 &  0.19 &  0.62 &  0.75 & -0.22 &  2.9 & -0.07 & 17.20 &   129 &   115 & \multicolumn{2}{c|}{ } \\
\hline
051228 & 22:40 &   ACE &  0.60 &  0.77 & -0.21 &  0.73 &  0.68 & -0.06 &  2.6 &  0.93 & 19.17 &    53 &    46 & \textbf{85} & 5 \\
\hline
051229 & 00:00  &  ACE &  0.59 & -0.17 & -0.79 &  0.72 & -0.12 & -0.69 & 13.4 &  0.56 &  7.28 &    54 &    98 & \textbf{83} & \textbf{74} \\
\hline
051229 & 01:20 &   C1  & -0.83 & -0.16 &  0.54 &  0.74 &  0.20 & -0.64 &  2.9 & -0.53 & 12.13 &   145 &    75 & \textbf{72} & \textbf{65} \\
\hline
051229 & 01:55  &  ACE &  0.18 & -0.55 & -0.82 & -0.16 &  0.54 &  0.83 &  3.4 &  0.07 & 15.24 &    79 &    45 & 1 & 1  \\
\hline
051229 & 02:28 &   ACE & \multicolumn{3}{c|}{ }  &  0.64 &  0.78 &  0.15 &   2.7 &   -0.50 &     16.80   & & &  \textbf{76} &  \textbf{73} \\
       & 02:28  &  C1  & -0.47 & -0.87 & -0.14 &  0.50 &  0.85 &  0.17 &  3.0 &  0.23 & 12.34 &   117 &    69 & \multicolumn{2}{c|}{ } \\
       & 02:28  &  C2 & \multicolumn{3}{c|}{ } &  0.42 &  0.89 &  0.18 &   1.8 &    0.08 &   21.17  & & & \multicolumn{2}{c|}{} \\
\hline \hline
060117 & 04:50  &  C3 & -0.16 & -0.75 & -0.64 & -0.21 &  0.82 &  0.53 &  1.4 &  0.99 & 26.55 &    99 &   107 &  \textbf{57} & \textbf{53}  \\
\hline
060126 & 21:22  &  C1 & \multicolumn{3}{c|}{ } &  0.38 &  0.92 &  0.06 &   2.4 &   -0.70 &     18.30   & & & \textbf{54} & \textbf{57} \\
       & 21:22  &  C2 & -0.72 & -0.70 &  0.03 &  0.28 &  0.96 &  0.07 &  4.4 & -1.07 & 11.74 &   135 &   154 & \multicolumn{2}{c|}{ } \\
       & 21:22  &  C3 & \multicolumn{3}{c|}{ } &  0.47  & 0.88  & 0.03  &  4.1  &  -0.52  &   12.70  & & & \multicolumn{2}{c|}{} \\
\hline
060128 & 05:56  &  C2 & \multicolumn{3}{c|}{ }   &   0.34 & -0.17 &  0.92 &   1.6  &   0.19  &   20.26  & & & \textbf{76} & \textbf{78} \\
       & 05:56  &  C3 &  0.03 & -0.14 &  0.99 &  0.51 &  0.85 &  0.12 &  2.2 &  0.60 & 14.34 &    88 &    55 & \multicolumn{2}{c|}{ } \\
\hline
060128 & 06:12 &   C1 & \multicolumn{3}{c|}{ } &  0.39 & -0.63 &  0.68 &   1.4  &   0.07  &   27.53 & & &  \textbf{72} &  29 \\
       & 06:12  &  C4 & -0.23 &  0.73 & -0.64 &  0.29 & -0.71 &  0.64 &  1.9 &  0.02 & 19.34 &   103 &   113 & \multicolumn{2}{c|}{ } \\
\hline
060128 & 07:24  &  ACE &  0.60 & -0.78 &  0.17 & -0.40 &  0.91 & -0.07 &  1.5 &  0.74 & 32.51 &    53 &    30 & 34 & 35 \\
\hline
060128 & 08:25  &  C1 & \multicolumn{3}{c|}{ }  &  -0.61 & -0.19 &  0.77 &   3.1 &   -0.71 &    12.54   & & & \textbf{46} & \textbf{45} \\
       & 08:25  &  C2 & \multicolumn{3}{c|}{ } & -0.56 & -0.18 &  0.81 &   2.3 &   -0.57 &    15.86  & & & \multicolumn{2}{c|}{} \\
       & 08:25  &  C3 & \multicolumn{3}{c|}{ } & -0.58 & -0.13 &  0.80 &   3.1 &   -0.76 &     12.58   & & & \multicolumn{2}{c|}{} \\
       & 08:25  &  C4 &  0.36 &  0.03 & -0.93 & -0.59 & -0.13 &  0.80 &  4.0 & -0.72 & 10.54 &    68 &    65 & \multicolumn{2}{c|}{ } \\
\hline
060128 & 13:25 &   C2 & -0.21 & -0.96 &  0.20 &  0.28 &  0.90 & -0.33 &  1.30 &  0.08 & 36.73 &   102 &    51 &  43 &  45 \\
\hline
060214 & 22:35 &   C3 & \multicolumn{3}{c|}{ }   &  -0.18 & -0.57 &  0.80 &   2.6 &   -0.17 &    11.70  & & & 45 & 24 \\
       & 22:35 &   C4 &  0.25 &  0.67 & -0.70 & -0.27 & -0.66 &  0.70 &  7.3 &  0.08 &  5.95 &    75 &    43 & \multicolumn{2}{c|}{ } \\
\hline
060215 & 23:29  &  C1 & \multicolumn{3}{c|}{ }  &   0.501 &  0.217 &  0.838 &   1.3 &   -0.34 &     24.82  & & &  \textbf{88} &  33 \\
       & 23:29 &   C2 & \multicolumn{3}{c|}{ } &  0.358  & 0.201  & 0.912  &  1.6  &  -0.26 &     16.32  & & & \multicolumn{2}{c|}{} \\
       & 23:29 &   C4 &  0.37 &  0.25 &  0.89 &  0.07 &  0.15 &  0.99 &  4.0 &  1.00 &  9.63 &    68 &   130 & \multicolumn{2}{c|}{ } \\
\hline
060221 & 01:47  &  ACE & -0.36 & -0.26 & -0.89 &  0.22 &  0.36 &  0.91 &  5.90 & -0.54 &  8.89 &   111 &    89 & 34 &  30 \\
       & 01:47  &  C1 & \multicolumn{3}{c|}{ } & -0.16  & 0.31  & 0.93 &   1.4 &    0.48 &   26.59   & & & \multicolumn{2}{c|}{} \\
\hline
060222 & 01:10 &   C3  &  0.39 &  0.76 &  0.52 &  0.48 &  0.72 &  0.50 &  2.30 &  0.24 & 23.05 &    66 &   107 & \textbf{80} & \textbf{84} \\
\hline
060223 & 04:14 &   C2 & \multicolumn{3}{c|}{ } &  0.62 & -0.26 & -0.74 &   2.2 &   -0.16 &    21.28  & & & \multicolumn{2}{c|}{} \\
       & 04:14  &  C3 &  0.62 & -0.39 & -0.68 &  0.59 & -0.41 & -0.69 &  2.3 &  0.12 & 20.69 &           51 &            7 &  \textbf{50} & \textbf{50} \\
\hline
060310 & 15:30 &   C3 &  0.96 & -0.26 &  0.14 &  0.99 & -0.05 &  0.12 &  4.1 & -0.35 & 13.06 &    17 &    78 & 41 & 44 \\
       & 15:30 &   C4 & \multicolumn{3}{c|}{ } &  0.94 & -0.24 &  0.23 &   3.9 &   -0.27 &    13.39  & & & \multicolumn{2}{c|}{} \\
\hline
060320 & 04:15 &   C1 & \multicolumn{3}{c|}{ }  &   0.55 & -0.21  & 0.81  &  2.7 &    1.00  &     11.08  & & & \textbf{64} &  \textbf{71} \\
       & 04:15 &   C2 & -0.19 &  0.17 & -0.97 &  0.36 & -0.15 &  0.92 &  5.2 &  0.67 &  8.18 &   101 &    90 & \multicolumn{2}{c|}{ } \\
       & 04:15 &   C4 & \multicolumn{3}{c|}{ }  &   0.68 & -0.41 &  0.61 &   2.0 &    1.47 &    15.12  & & & \multicolumn{2}{c|}{} \\
\hline
060322 & 07:58 &   C1 &  0.56 &  0.43 &  0.71 &  0.75 & -0.09 &  0.66 &  5.2 &  0.66 &  9.38 &    56 &   125 & 16 &  27 \\
       & 07:58 &   C2 & \multicolumn{3}{c|}{ }  &   0.74 & -0.10 &  0.67  & 4.6  &  0.69 &   10.09  & & & \multicolumn{2}{c|}{} \\
       & 07:58 &   C3 & \multicolumn{3}{c|}{ }  &  0.74 & -0.23 &  0.63 &  1.8  &  0.78 &  21.30  & & & \multicolumn{2}{c|}{} \\
\hline
060410 & 04:38 &   C4 &  0.69 &  0.14 &  0.71 &  0.54 &  0.23 &  0.81 &  1.3 &  0.98 & 29.94 &    46 &    42 &  \textbf{67} &  \textbf{66} \\   
\hline 
\bottomhline
\end{tabular}
\end{table*}

\setcounter{table}{1}

\begin{table*}[t]

\caption{Parameters of TD normal vectors: $\lambda_2/\lambda_3$ is the ratio of 2nd and 3rd eigenvalues, $B_{\mathrm{min}}$ is the smallest magnetic field component in minimum variance system, $\Delta\mathbf{n}$ is the error cone of minimum variance method, $\gamma$ is the angle between the Sun direction and TD normal, $\Delta\Phi$ is the direction change across the discontinuity and $\theta$ the angle between the bow shock normal and the $\mathbf{B}$ magnetic field vector. Boldface letter shows quasi-perpendicular conditions; the angles were calculated by scaling a model BS to the location of Cluster-1 and 3 spacecraft.} \label{tab:tddatac}
\vskip 4 mm
\centering

\begin{tabular}{|c|c|c|r@{,}r@{,}r|r@{,}r@{,}r|c|c|c|c|c|r@{,}r|}
\tophline
date & time & s/c & \multicolumn{3}{c|}{$\mathbf{n}_{\mathbf{B}_{u} \times \mathbf{B}_{d}}$} & \multicolumn{3}{c|}{$\mathbf{n}_{\mathrm{minvar}}$} & $\frac{\lambda_{2}}{\lambda_{3}}$ & $B_{\mathrm{min}}$ & $\Delta\mathbf{n}$ & $\gamma$ & $\Delta\Phi$ &  \multicolumn{2}{c|}{$\theta_{C1, C3}$} \\
(yymmdd) & (UT) & & \multicolumn{3}{c|}{} & \multicolumn{3}{c|}{} & & $\left(nT\right)$ & $\left({}^{o}\right)$ & $\left({}^{o}\right)$ & $\left({}^{o}\right)$ & \multicolumn{2}{c|}{$\left({}^{o}\right)$} \\
\middlehline 
060410 & 05:28 &   C1 & \multicolumn{3}{c|}{ } &   0.66 &  0.39 &  0.64 &   1.4 &   -0.96  &   35.69  & & &  41 &  43 \\  
       & 05:28  &  C3 &  0.53 &  0.56 &  0.64 &  0.49 &  0.59 &  0.64 &  1.7 &  0.15 & 27.49 &    58 &   116 & \multicolumn{2}{c|}{ } \\
       & 05:28  &  C4 & \multicolumn{3}{c|}{ } &   0.46 &  0.57 &  0.68 &   1.4 &   -0.55 &     39.17  & & & \multicolumn{2}{c|}{} \\
\hline
060410 & 07:53 &   C2 &  0.60 &  0.18 &  0.78 &  0.76 &  0.16 &  0.63 &  2.3 & -0.83 & 15.52 &    53 &    42 &  32 &  32 \\
       & 07:53 &   C4 & \multicolumn{3}{c|}{ } &   0.62 & -0.08  & 0.78 &   1.9  &  -1.01 &     20.03  & & & \multicolumn{2}{c|}{} \\
\hline
060410 & 08:30 &   C1 & \multicolumn{3}{c|}{ }  &   0.84 &  -0.01 &  0.55 &   1.1 &   -1.03 &    43.56 & & & 30 & 34 \\
       & 08:30 &   C2 & \multicolumn{3}{c|}{ }  &   0.80 &  0.07 &  0.60 &   3.7 &   -0.72 &     11.81   & & & \multicolumn{2}{c|}{} \\
       & 08:30 &   C3 &  0.62 &  0.39 &  0.68 &  0.76 &  0.13 &  0.64 &  4.9 & -0.62 &  9.79 &    51 &   114 & \multicolumn{2}{c|}{ } \\
       & 08:30 &   C4 & \multicolumn{3}{c|}{ }   &   0.76  & 0.11  & 0.64  &  4.7 &   -0.67 &     9.95  & & & \multicolumn{2}{c|}{} \\
\hline
060410 & 12:52 &   C1 & -0.15 & -0.12 &  0.98 & -0.57 &  0.05 &  0.82 &  2.6 &  1.59 & 15.89 &    98 &    17 & \textbf{81} & \textbf{81} \\
       & 12:52 &   C2 & \multicolumn{3}{c|}{ }   &   0.84 &  0.18 & -0.51  &  1.7 &   -0.85  &    24.58  & & & \multicolumn{2}{c|}{} \\
\hline
060416 & 12:38 &   C2 & -0.37 &  0.23 &  0.90 &  0.49 &  0.75 &  0.43 &  1.3 & -0.07 & 28.76 &   112 &     8 & 39 & \textbf{60} \\
\hline
060416 & 12:45 &   C1 & -0.06 & -0.69 & -0.72 & -0.09 &  0.77 &  0.63 &  5.4 & -0.75 & 11.20 &    93 &    36 & \multicolumn{2}{c|}{ } \\
       & 12:45 &   C2 & \multicolumn{3}{c|}{ }   &   0.09  & 0.73  & 0.68  &  2.1 &    0.46 &    22.20   & & & \multicolumn{2}{c|}{} \\
\hline
060416 & 13:24  &  C1 &  0.77 &  0.54 &  0.33 &  0.73 &  0.36 &  0.58 &  1.5 &  0.30 & 35.29 &    39 &    36 & \multicolumn{2}{c|}{ } \\
\hline
060416 & 15:56 &   C4 &  0.23 & -0.59 &  0.77 &  0.44 & -0.26 &  0.86 &  6.4 & -0.38 &  7.77 &    76 &    25 &  \textbf{48} & \textbf{47} \\
\hline
060416 & 16:29 &   C1 & \multicolumn{3}{c|}{ }   &   0.31 &  0.63  & 0.71  &  6.0 &    0.52 &     9.24   & & &  43 &  \textbf{54} \\
       & 16:29 &   C2 &  0.01 & -0.81 & -0.58 &  0.14 &  0.75 &  0.65 & 14.4 &  0.30 &  5.63 &    89 &   126 & \multicolumn{2}{c|}{ } \\
\hline
060416 & 16:40  &  ACE & -0.78 & -0.37 & -0.51 &  0.85 &  0.12 &  0.52 &  1.7 &  0.44 & 20.40 &   140 &   109 &  43 &  \textbf{54} \\
       & 16:40  &  C1 & \multicolumn{3}{c|}{ }   &  0.83 & -0.08 & -0.55 &   1.4 &   -0.11 &     29.50   & & & \multicolumn{2}{c|}{} \\
\hline
060416 & 18:33 &   ACE &  0.30 & -0.09 & -0.95 & -0.48 & -0.17 &  0.86 &  3.5 & -0.35 & 12.03 &    72 &    22 & 44 & \textbf{49} \\
\hline
060416 & 20:01 &   C3 & -0.05 &  0.29 &  0.96 &  0.17 &  0.44 &  0.88 &  2.0 & -0.03 & 14.97 &    92 &    10 & \textbf{49} & \textbf{48} \\
       & 20:01 &   C4 & \multicolumn{3}{c|}{ }  &   0.29  & 0.43  & 0.85 &   1.5 &   -0.20 &    21.62  & & & \multicolumn{2}{c|}{} \\
\hline
\hline
070104 & 03:54 &   C2  &  0.83 & -0.33 &  0.45 &  0.25 &  0.93 &  0.29 &  1.3 & -0.50 & 32.78 &  33 &   27 &  44 &  44 \\
\hline
070104 & 04:38 &   ACE & -0.63 &  0.64 & -0.45 &  0.68 & -0.72 &  0.13 & 10.3 & -1.48 &  7.73 &  128 &   26 &  31 &  31 \\
\hline
070104 & 05:08 &   ACE & \multicolumn{3}{c|}{} &   0.60 & -0.54 & 0.58 &  2.2  & -0.03 &  19.08 & & & \multicolumn{2}{c|}{} \\
       & 05:08 &   C1 & \multicolumn{3}{c|}{} &   0.71 & -0.61 & 0.35 &  2.9  & -0.16 &   14.68 & & & \multicolumn{2}{c|}{} \\
       & 05:08 &   C2 & \multicolumn{3}{c|}{} &    0.57 & -0.68 &  0.46 &  2.1  &  0.12 &  19.78 & & & \multicolumn{2}{c|}{} \\
       & 05:08 &   C3 &  0.64 & -0.64 &  0.43 &  0.68 & -0.62 &  0.40 &  3.0 & -0.15 & 13.51 &   50 &  113 & \multicolumn{2}{c|}{} \\
       & 05:08 &   C4 & \multicolumn{3}{c|}{} &    0.58 & -0.67 & 0.46 &  2.6  &  0.10 &  16.49 & & & \multicolumn{2}{c|}{} \\
\hline
070104 & 06:20 &   ACE &  0.62 &  0.04 &  0.78 &  0.61 &  0.05 &  0.79 &  7.1 &  0.04 &  8.63 &   51 &   96 & 23 & 23 \\
       & 06:20 &   C1 & \multicolumn{3}{c|}{} &   0.73 &  0.30 &  0.62 &   1.8 &    -0.05 &   20.18 & & & \multicolumn{2}{c|}{} \\
\hline
070106 & 16:10 &   C1 &  0.24 &  0.44 &  0.86 &  0.21 &  0.76 &  0.61 &  1.7 &  0.66 & 20.21 &   75 &    8 & 10 & 5 \\
\hline
070108 & 11:25 &   ACE & \multicolumn{3}{c|}{} &  -0.51 & 0.86 & -0.01 &  2.2  &  0.03 &  17.13 & & & \textbf{82} & \textbf{74} \\
       & 11:25 &   C1 & -0.62 &  0.78 &  0.01 & -0.66 &  0.75 &  0.02 &  2.4 & -0.04 & 12.24 &  128 &  129 & \multicolumn{2}{c|}{} \\
       & 11:25 &   C2 & \multicolumn{3}{c|}{} &   -0.67 & 0.74 & 0.06 &  2.1   & -0.05  & 15.71 & & & \multicolumn{2}{c|}{} \\
       & 11:25 &   C3 & \multicolumn{3}{c|}{} &   -0.50 & 0.86 & -0.06  & 1.2  &  0.04 &  37.42 & & & \multicolumn{2}{c|}{} \\
       & 11:25 &   C4 & \multicolumn{3}{c|}{} &    0.50 & -0.86 &  0.05 &  1.3   & -0.08 &  30.75 & & & \multicolumn{2}{c|}{} \\
\hline
070116 & 09:41 &   C1 &  0.36 &  0.29 &  0.88 &  0.36 &  0.17 &  0.92 &  2.3 & -0.37 & 14.57 &   68 &    6 &  24 &  22 \\
\hline
070116 & 10:00 &   ACE &  0.98 & -0.11 &  0.15 &  0.94 & -0.29 & -0.17 &  2.2 &  0.25 & 17.96 &   10 &  160 &  21 &  16 \\

\hline
070116 & 10:50 &   C1 & \multicolumn{3}{c|}{} &   0.74 & -0.64 & 0.20 &  3.9   & -0.12 &  9.57 & & &  28 &  28 \\
       & 10:50 &   C2 & \multicolumn{3}{c|}{} &    0.73 & -0.63 & 0.26 &  3.0   & -0.11 &  11.49 & & & \multicolumn{2}{c|}{} \\
       & 10:50 &   C3 & -0.73 &  0.61 & -0.31 &  0.68 & -0.61 &  0.40 &  4.4 &  0.22 &  8.72 &  136 &   43 & \multicolumn{2}{c|}{} \\

070116 & 10:50 &   C4 & \multicolumn{3}{c|}{} &    0.69 & -0.62 &  0.37  & 4.4 &   0.16  & 8.70 & & & \multicolumn{2}{c|}{} \\
\hline
070117 & 16:40 &   ACE & -0.49 & -0.74 & -0.46 &  0.55 &  0.55 &  0.63 &  1.9 & -0.80 & 24.59 &  119 &   82 & 9 &  12 \\
\hline
070118 & 07:52 &   C3 &  0.50 &  0.85 &  0.14 &  0.45 &  0.88 &  0.18 &  1.5 &  0.30 & 25.86 &   59 &   57 & 28 &  24 \\   
070118 & 07:52 &   C4 & \multicolumn{3}{c|}{} &    0.51 &  0.85 & 0.12 &  1.1   & -0.19  & 48.52 & & & \multicolumn{2}{c|}{} \\
\hline
070118 & 09:38 &   C2 &  0.85 & -0.25 &  0.46 &  0.87 & -0.29 &  0.39 &  1.5 & -0.33 & 22.97 &   31 &   55 & 26 & 17 \\
\hline
070118 & 09:44 &   C1 &  0.76 &  0.27 &  0.58 &  0.76 &  0.24 &  0.60 &  2.3 &  0.04 & 15.52 &   40 &  137 & \multicolumn{2}{c|}{} \\
\hline
\bottomhline
\end{tabular}
\end{table*}

\setcounter{table}{1}

\begin{table*}[t]

\caption{Parameters of TD normal vectors: $\lambda_2/\lambda_3$ is the ratio of 2nd and 3rd eigenvalues, $B_{\mathrm{min}}$ is the smallest magnetic field component in minimum variance system, $\Delta\mathbf{n}$ is the error cone of minimum variance method, $\gamma$ is the angle between the Sun direction and TD normal, $\Delta\Phi$ is the direction change across the discontinuity and $\theta$ the angle between the bow shock normal and the $\mathbf{B}$ magnetic field vector. Boldface letter shows quasi-perpendicular conditions; the angles were calculated by scaling a model BS to the location of Cluster-1 and 3 spacecraft.} \label{tab:tddatad}
\vskip 4 mm
\centering

\begin{tabular}{|c|c|c|r@{,}r@{,}r|r@{,}r@{,}r|c|c|c|c|c|r@{,}r|}
\tophline
date & time & s/c & \multicolumn{3}{c|}{$\mathbf{n}_{\mathbf{B}_{u} \times \mathbf{B}_{d}}$} & \multicolumn{3}{c|}{$\mathbf{n}_{\mathrm{minvar}}$} & $\frac{\lambda_{2}}{\lambda_{3}}$ & $B_{\mathrm{min}}$ & $\Delta\mathbf{n}$ & $\gamma$ & $\Delta\Phi$ &  \multicolumn{2}{c|}{$\theta_{C1, C3}$} \\
(yymmdd) & (UT) & & \multicolumn{3}{c|}{} & \multicolumn{3}{c|}{} & & $\left(nT\right)$ & $\left({}^{o}\right)$ & $\left({}^{o}\right)$ & $\left({}^{o}\right)$ & \multicolumn{2}{c|}{$\left({}^{o}\right)$} \\
\middlehline 
070118 & 12:15 &   C1 & -0.80 & -0.10 &  0.59 & -0.84 &  0.08 & -0.53 &  2.9 & -0.18 & 12.16 &  143 &   87 &  38 &  9 \\
\hline
070118 & 14:36 &   C1 & \multicolumn{3}{c|}{} &    0.07 &  1.00 & -0.04 &  2.2  &  0.03 &  13.11 & & & 36 &  14 \\
       & 14:36 &   C2 & \multicolumn{3}{c|}{} &    0.04 &  1.00 & -0.02 &  2.7  &  0.05 &  10.83 & & & \multicolumn{2}{c|}{} \\
       & 14:36 &   C3 & -0.02 & -1.00 &  0.02 &  0.03 &  1.00 & -0.15 &  2.9 & -0.60 & 10.69 &   91 &   28 & \multicolumn{2}{c|}{} \\
\hline
070118 & 19:35 &   C1 & \multicolumn{3}{c|}{} &    0.52 & 0.79 & -0.33 &  2.7   & -0.18 &  12.26 & & & \textbf{87} & \textbf{87} \\
       & 19:35 &   C2 & \multicolumn{3}{c|}{} &    0.55 & 0.77 & -0.31 & 2.1  &  0.01 &  14.83 & & & \multicolumn{2}{c|}{} \\
       & 19:35 &   C3 & \multicolumn{3}{c|}{} &    0.41 &  0.80 & -0.44 &  2.4 &   0.23 &  13.41 & & & \multicolumn{2}{c|}{} \\
       & 19:35 &   C4 &  0.50 &  0.77 & -0.39 &  0.46 &  0.80 & -0.40 &  3.3 &  0.10 & 10.51 &   60 &   81 & \multicolumn{2}{c|}{} \\
\hline
070120 & 18:20 &   C1 & \multicolumn{3}{c|}{} &    0.69 & -0.40 & 0.60 &  7.2   & -0.42 &  6.28 & & & \multicolumn{2}{c|}{} \\
       & 18:20 &   C2 &  0.51 & -0.42 &  0.75 &  0.60 & -0.41 &  0.69 & 10.3 & -0.31 &  5.19 &   59 &  100 & \multicolumn{2}{c|}{} \\
       & 18:20 &   C3 & \multicolumn{3}{c|}{} &    0.61 & -0.40 &  0.68  & 8.3   & -0.40 &  5.89 & & & \multicolumn{2}{c|}{} \\
       & 18:20 &   C4 & \multicolumn{3}{c|}{} &    0.63 & -0.41 &  0.66 &  7.6   & -0.51 &  6.14 & & & \multicolumn{2}{c|}{} \\
\hline
070130 & 16:47 &   ACE &  0.67 & -0.58 &  0.46 &  0.83 & -0.46 &  0.31 &  1.8 & -0.80 & 28.51 &   47 &   26 &  23 & 17 \\
\hline
070201 & 06:49 &   C1 &  0.24 & -0.33 &  0.91 &  0.26 & -0.33 &  0.91 &  2.2 & -0.07 & 15.73 &   75 &   71 &  41 & 44  \\
       & 06:49 &   C2 & \multicolumn{3}{c|}{} &   -0.11 & -0.39 & 0.91 &  1.0  &  0.57 &  104.85& & & \multicolumn{2}{c|}{} \\
       & 06:49 &   C3 & \multicolumn{3}{c|}{} &   -0.20 & -0.47 & 0.86 &  1.3  &  0.87 &  32.17 & & & \multicolumn{2}{c|}{} \\
       & 06:49 &   C4 & \multicolumn{3}{c|}{} &   -0.11 & -0.47 & 0.88 &  1.2  &  0.66 &  41.45 & & & \multicolumn{2}{c|}{} \\ 
\hline
070201 & 22:08 &   C1 &  0.30 & -0.89 &  0.33 & -0.35 &  0.88 & -0.31 &  3.0 &  0.05 & 12.58 &   72 &   37 &  \textbf{64} & \textbf{61} \\
       & 22:08 &   C2 & \multicolumn{3}{c|}{} &   -0.38 & 0.90 & -0.21 &  2.6  &  0.65 &  13.92 & & & \multicolumn{2}{c|}{} \\
       & 22:08 &   C3 & \multicolumn{3}{c|}{} &   -0.43 & 0.86 & -0.29 &  2.7  &  0.16 &  13.66 & & & \multicolumn{2}{c|}{} \\
       & 22:08 &   C4 & \multicolumn{3}{c|}{} &   -0.42 & 0.85 & -0.30 &  1.9  &  0.16 &  18.62 & & & \multicolumn{2}{c|}{} \\
\hline
070201 & 22:17 &   ACE &  0.22 & -0.89 &  0.41 &  0.15 & -0.89 &  0.44 & 16.9 &  0.16 &  6.57 &   77 &   64 & \textbf{57} &  43 \\
       & 22:17 &   C3 & \multicolumn{3}{c|}{} &   -0.23 &  0.84 & -0.48 &  2.1  &  0.13 &  21.55 & & & \multicolumn{2}{c|}{} \\
\hline
070202 & 01:31 &   ACE & \multicolumn{3}{c|}{} &  -0.42 & 0.91 &  0.02 &  2.5  &  0.08 &  15.51 & & &  \textbf{83} &  \textbf{83} \\
       & 01:31 &   C1 &  0.53 & -0.84 & -0.10 & -0.57 &  0.80 &  0.18 &  4.0 &  0.29 &  9.43 &   58 &   25 & \multicolumn{2}{c|}{} \\
       & 01:31 &   C2 & \multicolumn{3}{c|}{} &   -0.44 &  0.89 & 0.05 &  3.0   & -0.26  & 12.34 & & & \multicolumn{2}{c|}{} \\
\hline
070209 & 02:16 &   C1 & -0.96 & -0.25 &  0.14 &  0.94 &  0.29 & -0.16 &  5.8 & -0.13 &  8.62 &  163 &   89 & \textbf{72} &  \textbf{63} \\
       & 02:16 &   C2 & \multicolumn{3}{c|}{} &    0.92 &  0.35 & -0.19 &  5.8   & -0.16 &  8.68 & & & \multicolumn{2}{c|}{} \\
\hline
070215 & 01:35 &   C2 &  0.31 & -0.05 & -0.95 & -0.38 &  0.14 &  0.91 &  2.2 &  0.31 & 14.35 &   71 &   55 & 25 & 28 \\
\hline
070215 & 02:31 &   ACE & \multicolumn{3}{c|}{} &   0.66 & -0.21 &  0.73 &  3.0 &   0.01 &  13.32 & & & 25 &  28 \\
       & 02:31 &   C1 &  0.72 & -0.33 &  0.61 &  0.70 & -0.33 &  0.63 &  5.9 &  0.06 &  6.73 &   43 &  102 & \multicolumn{2}{c|}{} \\
       & 02:31 &   C3 & \multicolumn{3}{c|}{} &    0.55 & -0.35 &  0.75 &  1.6   & -0.05 &  19.96 & & & \multicolumn{2}{c|}{} \\
       & 02:31 &   C4 & \multicolumn{3}{c|}{} &    0.52 & -0.37 & 0.77 &  1.7 &   0.10 &  15.51 & & & \multicolumn{2}{c|}{} \\
\hline
070215 & 02:50 &   ACE & \multicolumn{3}{c|}{} &  -0.40 & -0.22 & 0.89 &  3.7   & -0.11 &  13.76 & & &  25 &  28 \\
       & 02:50 &   C1 & \multicolumn{3}{c|}{} &   0.63 & -0.06 & 0.77 &  1.2   & -0.14 &  26.49 & & & \multicolumn{2}{c|}{} \\
       & 02:50 &   C2 &  0.14 &  0.19 & -0.97 &  0.80 &  0.30 & -0.53 &  4.5 & -1.09 &  5.53 &   82 &  138 & \multicolumn{2}{c|}{} \\
       & 02:50 &   C3 & \multicolumn{3}{c|}{} &    0.66 & -0.17 &  0.73 &  1.6  &  0.00 &  16.97  & & & \multicolumn{2}{c|}{} \\
       & 02:50 &   C4 & \multicolumn{3}{c|}{} &    0.68 & -0.17  & 0.71 &  1.4  &  0.00 &  19.53 & & & \multicolumn{2}{c|}{} \\
\hline
070215 & 03:13 &   C1 &  0.43 & -0.27 & -0.86 &  0.70 & -0.54 & -0.47 &  1.6 & -0.27 & 24.17 &   64 &    8 &  27 & 29 \\
\hline
070215 & 03:56 &   C4 &  0.43 &  0.31 &  0.85 &  0.66 &  0.35 &  0.66 &  9.1 & -1.25 &  6.78 &   64 &  100 &  25 &  25 \\
\hline
070215 & 04:00 &   C1 & -0.06 & -0.71 & -0.70 &  0.05 &  0.75 &  0.65 &  1.9 & -0.13 & 20.13 &   93 &   20 & \multicolumn{2}{c|}{} \\
\hline
070215 & 08:45 &   C1 &  0.82 &  0.41 &  0.40 &  0.79 &  0.44 &  0.42 &  6.5 & -0.08 &  6.81 &   35 &   66 &  10 &  25 \\
       & 08:45 &   C2 & \multicolumn{3}{c|}{} &    0.84 &  0.17 & 0.51 &  2.3  &  0.10 &  16.65 & & & \multicolumn{2}{c|}{} \\
       & 08:45 &   C3 & \multicolumn{3}{c|}{} &    0.75 & 0.26 & 0.61 &  1.6  &  0.18 &  26.22 & & & \multicolumn{2}{c|}{} \\
       & 08:45 &   C4 & \multicolumn{3}{c|}{} &    0.70 & 0.20 & 0.68 &  1.6  &  0.29 &  25.26 & & & \multicolumn{2}{c|}{} \\
\hline
070215 & 15:15 &   C3 &  0.45 & -0.53 &  0.72 &  0.49 & -0.53 &  0.69 &  3.4 & -0.14 & 11.26 &   63 &   74 &  44 & \textbf{53} \\
070215 & 15:15 &   C4 & \multicolumn{3}{c|}{} &    0.48 & -0.51 & 0.71 &  2.7   & -0.08 &  13.36 & & & \multicolumn{2}{c|}{} \\
\hline
070215  & 22:08 &   C2 & \multicolumn{3}{c|}{} &   -0.38 & 0.90 & -0.21 &  2.6  &  0.65 &  13.92 & & & \multicolumn{2}{c|}{} \\
\hline
070301 & 04:56 &   C2 &  0.75 & -0.66 & -0.12 &  0.74 & -0.67 & -0.02 &  1.9 & -0.10 & 15.83 &   41 &   62 &  35 &  35 \\ 
\hline
\bottomhline
\end{tabular}
\end{table*}

\setcounter{table}{1}

\begin{table*}[t]

\caption{Parameters of TD normal vectors: $\lambda_2/\lambda_3$ is the ratio of 2nd and 3rd eigenvalues, $B_{\mathrm{min}}$ is the smallest magnetic field component in minimum variance system, $\Delta\mathbf{n}$ is the error cone of minimum variance method, $\gamma$ is the angle between the Sun direction and TD normal, $\Delta\Phi$ is the direction change across the discontinuity and $\theta$ the angle between the bow shock normal and the $\mathbf{B}$ magnetic field vector. Boldface letter shows quasi-perpendicular conditions; the angles were calculated by scaling a model BS to the location of Cluster-1 and 3 spacecraft.} \label{tab:tddatae}
\vskip 4 mm
\centering

\begin{tabular}{|c|c|c|r@{,}r@{,}r|r@{,}r@{,}r|c|c|c|c|c|r@{,}r|}
\tophline
date & time & s/c & \multicolumn{3}{c|}{$\mathbf{n}_{\mathbf{B}_{u} \times \mathbf{B}_{d}}$} & \multicolumn{3}{c|}{$\mathbf{n}_{\mathrm{minvar}}$} & $\frac{\lambda_{2}}{\lambda_{3}}$ & $B_{\mathrm{min}}$ & $\Delta\mathbf{n}$ & $\gamma$ & $\Delta\Phi$ &  \multicolumn{2}{c|}{$\theta_{C1, C3}$} \\
(yymmdd) & (UT) & & \multicolumn{3}{c|}{} & \multicolumn{3}{c|}{} & & $\left(nT\right)$ & $\left({}^{o}\right)$ & $\left({}^{o}\right)$ & $\left({}^{o}\right)$ & \multicolumn{2}{c|}{$\left({}^{o}\right)$} \\
\middlehline 
070301 & 04:56 &   C3 & \multicolumn{3}{c|}{} &    0.64 & -0.77 & -0.06 &  1.2   & -0.05 &  32.86 & & & \multicolumn{2}{c|}{} \\ 
       & 04:56 &   C4 & \multicolumn{3}{c|}{} &    0.64 & -0.76 & -0.11 &  1.2  &  0.07 &  35.48 & & & \multicolumn{2}{c|}{} \\
\hline
070301 & 07:11 &   C1 & \multicolumn{3}{c|}{} &   0.97 & -0.13 &  0.20 &   2.2   & -0.15 &  12.07 & & &  20 &  21 \\
       & 07:11 &   C2 & -0.93 &  0.16 & -0.33 &  0.93 & -0.15 &  0.35 &  2.9 &  0.03 &  9.77 &  158 &   99 & \multicolumn{2}{c|}{} \\
       & 07:11 &   C3 & \multicolumn{3}{c|}{} &    0.92 & -0.18 &  0.34 &  2.5   & -0.08 &  10.93 & & & \multicolumn{2}{c|}{} \\
\hline
070301 & 09:43 &   C1 & \multicolumn{3}{c|}{} &   0.47 & -0.80 & -0.37 &  1.7 &   0.21 &  15.12 & & &  \textbf{87} & \textbf{89} \\
       & 09:43 &   C2 &  0.72 & -0.69 & -0.03 &  0.78 & -0.60 & -0.15 &  2.0 & -0.34 & 12.93 &   43 &   17 & \multicolumn{2}{c|}{} \\
\hline
070301 & 10:30 &   C1 & -0.61 &  0.64 & -0.46 & -0.56 &  0.79 & -0.26 &  1.9 & -0.29 & 15.42 &  127 &    8 &  16 &  23 \\
\hline
070302 & 02:03 &   C1 &  0.71 &  0.69 &  0.12 &  0.80 &  0.57 &  0.18 &  2.1 & -0.17 & 12.71 &   44 &    8 &  \textbf{47} & \textbf{51} \\
\hline
070313 & 05:37 &   C2 &  0.60 &  0.52 & -0.61 &  0.61 &  0.52 & -0.60 &  5.2 & -0.05 &  7.48 &   53 &   92 & 24 & 23 \\
       & 05:37 &   C3 & \multicolumn{3}{c|}{} &    0.67 & 0.52 & -0.53 &  3.2   & -0.53 &   11.53 & & & \multicolumn{2}{c|}{} \\
       & 05:37 &   C4 & \multicolumn{3}{c|}{} &    0.69 & 0.49 & -0.52 &  3.8   & -0.48 &  10.35 & & & \multicolumn{2}{c|}{} \\
\hline
070314 & 07:54 &   ACE &  0.37 &  0.32 &  0.87 &  0.38 &  0.29 &  0.88 &  2.9 & -0.14 & 15.87 &   68 &   68 &  39 &  39 \\
\hline
070314 & 08:37 &   C1 & \multicolumn{3}{c|}{} &   0.74 & 0.43 &  0.52 &  1.8  &  0.27 &  30.34 & & & 45 &  39 \\
       & 08:37 &   C3 &  0.76 &  0.31 &  0.57 &  0.77 &  0.00 &  0.64 &  4.0 &  0.08 & 15.22 &   40 &   17 & \multicolumn{2}{c|}{} \\
       & 08:37 &   C4 & \multicolumn{3}{c|}{} &    0.70 & 0.20 &  0.68 &  2.9 &   0.28 &  18.32 & & & \multicolumn{2}{c|}{} \\
\hline
070314 & 12:51 &   ACE & -0.17 &  0.21 & -0.96 &  0.30 & -0.50 &  0.81 &  2.0 & -0.18 & 22.90 &   99 &   26 &  \textbf{51} &  \textbf{51} \\
\hline
070314 & 15:52 &   ACE & -0.85 & -0.46 & -0.26 &  0.86 &  0.52 & -0.03 &  1.9 &  0.16 & 21.58 &  147 &  163 &  9 & 9 \\
\hline
070315 & 12:15 &   ACE &  0.40 & -0.23 &  0.89 &  0.48 & -0.08 &  0.87 &  2.2 &  0.32 & 20.23 &   66 &  109 & \textbf{58} & \textbf{58} \\
       & 12:15 &   C1 & \multicolumn{3}{c|}{} &   0.37 & -0.40 &  0.84 &  1.5   & -0.07 &  22.67 & & & \multicolumn{2}{c|}{} \\
\hline
070316 & 18:14 &   C1 & -0.15 & -0.98 &  0.12 &  0.17 &  0.97 & -0.18 &  2.0 & -0.22 & 14.41 &   98 &   12 &  40 &  39 \\
\hline
070316 & 19:57 &   C1 & \multicolumn{3}{c|}{} &   0.64 & -0.61 & 0.46 &  1.6   & -0.08 &  17.19 & & & \textbf{51} & \textbf{53} \\
       & 19:57 &   C2 & \multicolumn{3}{c|}{} &    0.56 & -0.53 & 0.63 &   1.6   & -0.02 &  16.57 & & & \multicolumn{2}{c|}{} \\
       & 19:57 &   C3 & \multicolumn{3}{c|}{} &    0.41 & -0.51 &  0.75 &   1.2  &  0.13 &  30.65 & & & \multicolumn{2}{c|}{} \\
       & 19:57 &   C4 &  0.37 & -0.39 &  0.84 &  0.31 & -0.50 &  0.81 &  1.7 &  0.17 & 16.15 &   68 &   38 & \multicolumn{2}{c|}{} \\
\hline
070319 & 03:47 &   C2 & \multicolumn{3}{c|}{} &    0.83 &  0.03 & 0.56 &  1.1   & -0.04 &  48.73 & & & \textbf{49} & 30 \\
070319 & 03:47 &   C4 &  0.56 &  0.14 &  0.82 &  0.20 &  0.19 &  0.96 &  2.1 &  0.30 & 14.07 &   56 &  144 & \multicolumn{2}{c|}{} \\
\hline
070319 & 04:28 &   C1 & \multicolumn{3}{c|}{} &   0.50 &  0.42 & 0.76 &  2.1   & -0.14 &  13.22   & & &  16 & 13 \\
070319 & 04:28 &   C2 & \multicolumn{3}{c|}{} &    0.32 &  0.40 &  0.86 &  1.9 &   0.02 &  15.32 & & & \multicolumn{2}{c|}{} \\
070319 & 04:28 &   C3 & -0.24 & -0.39 & -0.89 &  0.20 &  0.38 &  0.90 &  2.3 &  0.06 & 12.42 &  104 &   96 & \multicolumn{2}{c|}{} \\
\hline
070328 & 13:41 &   C3 &  0.90 &  0.31 &  0.29 &  0.90 &  0.30 &  0.30 &  1.5 & -0.01 & 20.23 &   25 &   47 & 9 & \textbf{88} \\
       & 13:41 &   C4 & \multicolumn{3}{c|}{} &    0.86 & 0.38 &  0.34 &  1.5  &  0.11 &  20.58 & & & \multicolumn{2}{c|}{} \\
\hline
070328 & 15:22 &   C2 &  0.23 & -0.93 & -0.28 & -0.24 &  0.96 & -0.15 &  1.9 & -0.06 & 24.12 &   76 &    4 &  35 & 30 \\
\hline
070328 & 16:08 &   C2 &  0.04 &  1.00 &  0.05 & -0.09 &  0.95 &  0.29 &  1.4 & -0.30 & 23.79 &   87 &   13 & 36 & \textbf{47} \\
\hline
070328 & 16:51 &   C1 & \multicolumn{3}{c|}{} &   0.63 & -0.22 & -0.74 &  1.7  &  0.01 &  16.24 & & & \textbf{66} & \textbf{71} \\
       & 16:51 &   C2 & \multicolumn{3}{c|}{} &    0.51 & -0.24 & -0.82  & 2.0  &  0.00  & 13.11 & & & \multicolumn{2}{c|}{} \\
       & 16:51 &   C4 &  0.05 & -0.15 & -0.99 & -0.50 &  0.28 &  0.82 &  2.4 & -0.34 & 11.23 &   86 &   16 & \multicolumn{2}{c|}{} \\
\hline
070429 & 20:41 &   C2 & -0.79 & -0.55 &  0.29 &  0.79 &  0.54 & -0.29 &  2.6 &  0.04 & 12.24 &  141 &   68 &  39 & 40 \\
\hline
070429 & 21:00 &   C1 & -0.79 & -0.48 &  0.37 &  0.72 &  0.35 & -0.60 &  1.4 &  0.66 & 18.29 &  142 &   27 &  \textbf{51} & \textbf{52} \\
\hline
070429 & 22:06 &   C1 & -0.37 & -0.90 & -0.23 &  0.33 &  0.85 &  0.41 &  1.2 & -0.52 & 27.17 &  111 &   44 & 39 &  40 \\
\hline
070429 & 23:02 &   ACE & -0.77 &  0.14 & -0.63 &  0.72 & -0.11 &  0.69 &  4.9 & -0.18 & 10.03 &  140 &  115 & \multicolumn{2}{c|}{} \\
       & 23:02 &   C1 & \multicolumn{3}{c|}{} &   0.66 & -0.04 & 0.75 &  1.8  &  0.14 &  14.56 & & & \multicolumn{2}{c|}{} \\
       & 23:02 &   C2 & \multicolumn{3}{c|}{} &    0.72 & -0.07 &  0.69 &  1.9 &   0.26 &  13.85 & & & \multicolumn{2}{c|}{} \\
       & 23:02 &   C3 & \multicolumn{3}{c|}{} &    0.65 & -0.03 & 0.76 &  2.3   & -0.09 &  11.31  & & & \multicolumn{2}{c|}{} \\
       & 23:02 &   C4 & \multicolumn{3}{c|}{} &    0.71 & -0.03 & 0.70 &  1.9   & -0.04 &  15.05  & & & \multicolumn{2}{c|}{} \\
\hline
070430 & 02:02 &   C1 & -0.23 & -0.46 & -0.86 &  0.65 &  0.40 &  0.64 &  6.3 &  0.65 &  5.92 &  103 &  146 & \multicolumn{2}{c|}{} \\
       & 02:02 &   C2 & \multicolumn{3}{c|}{} &    0.67 & 0.40 &  0.62 &  5.6   & 0.70 &  6.39 & & & \multicolumn{2}{c|}{} \\
       & 02:02 &   C3 & \multicolumn{3}{c|}{} &    0.68 &  0.40 & 0.61 &  5.5  &  0.48 &  6.50 & & & \multicolumn{2}{c|}{} \\
       & 02:02 &   C4 & \multicolumn{3}{c|}{} &    0.67 & 0.41 & 0.62 &  5.0   & 0.26 &  6.86 & & & \multicolumn{2}{c|}{} \\
\hline
\bottomhline
\end{tabular}
\end{table*}

The two angles (the $\gamma$ and the $\Delta\Phi$) mentioned before are considered to be very important in the formation of HFA events. We are able to measure these angles and thus to compare the results of measurements with the predictions of earlier simulations. Unfortunately, triangulation techniques can not be used to determine these angles because of strong magnetic field fluctuations. Thus the direction of the TD normal vector was determined by the cross-product method and minimum-variance techniques using Cluster FGM \citep{balogh01:_clust_magnet_field_inves} and ACE MAG measurements data \citep{smith98:_ace_magnet_field_exper}. The temporal resolution of FGM data series were 1\,s and MAG's resolution was 16\,s. We accepted the result of minimum variance method if the cross product method did not differ by more than 15\degree\ and the ratio of second and third eigenvalues were equal to or larger than 2.0 (Tab.~\ref{tab:tddataa}) (For a more detailed description of the method see \citet{facsko08:_statis_study_of_hot_flow, facsko08:_clust_hot_flow_anomal_obser}). It turns out that the minimum variance method can mostly be used at low magnetic field variation. This method is very difficult and almost impossible to use in the HFA cavity and in SLAMS (Short Large Amplitude Magnetic Structures) mostly coupled to quasi-parallel regions \citep{schwartz91:_quasi_paral_shock_patch_of}. Many HFAs were embedded into SLAMS and so we were able to use the minimum variance method with good accuracy only in a few cases. Beside of this feature of the method we have found more HFAs at the quasi-parallel region ($\sim$66\,\%). (See Tab.~\ref{tab:tddataa}) The local bow shock normals were calculated by scaling a model bow shock to the spacecraft location as in \citet{schwartz00:_condit} and we used the upstream magnetic field upstream of the HFA to calculate the angle of the shock-normal and the magnetic field vector. This might confirm previous results: the conditions were quasi-parallel at least on one side of the TD previously \citep[See][]{onsager91:_inter_reflec, thomsen93:_obser_test_hot_flow_abnom, kecskemety06:_distr_rapid_clust} and current simulations expect the HFAs to appear where the quasi-parallel condition turns to quasi-perpendicular \citep{omidi07:_format}. We used the same conditions for HFA observation and determination in 2003 \citep{kecskemety06:_distr_rapid_clust, facsko08:_statis_study_of_hot_flow}, 2006 and 2007 \citep{facsko08:_clust_hot_flow_anomal_obser}, however this effect was very strong in 2007 and it was also noticeable in 2003 and 2006. 

\begin{figure}[ht]
\centering
\epsfig{file=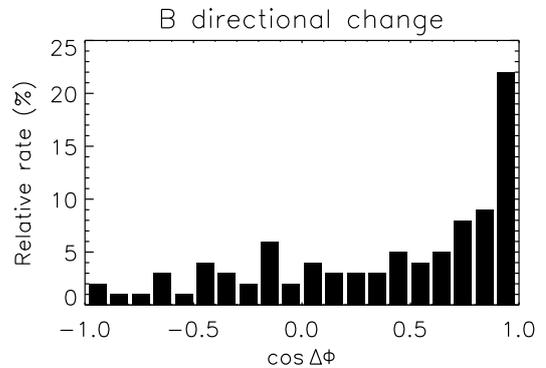, width=200pt}
\caption{Distribution of $\cos \left(\Delta\Phi\right)$ where $\Delta\Phi$ is the angle of magnetic field directional change at the discontinuity.}
\label{fig:dphidist}
\end{figure}

\begin{figure}[hb]
\begin{center}
\epsfig{file=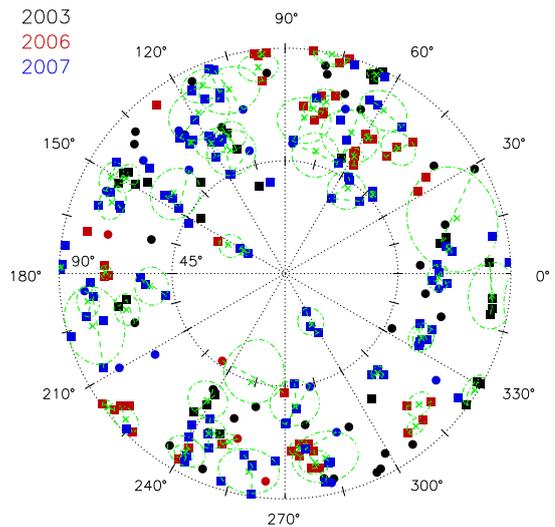, width=200pt}
\end{center}
\caption{Polar plot of the direction of the normal vectors of TDs. The azimuthal angle is measured between the GSE y direction and the projection of the normal vector onto the GSE yz plane. The distance from the center is the $\gamma$ angle as determined by the cross-product method. The TD normal vector is in a special polar coordinate system in which we measure the $\gamma$ angle from the center, and where the azimuth is the angle of GSE y and the projection of normal vector to GSE yz plane. The regions surrounded by dashed lines are the projection of error cones around the average normal vector marked by ``X''. Circles and squares symbolize ACE and Cluster data, respectively. The black, red and blue symbols present events observed in 2003, 2006 and 2007, respectively.}
\label{fig:cone}
\end{figure}

The $\Delta\Phi$ and $\gamma$ distributions differ from the typical distributions associated with discontinuities in the solar wind. The $\Delta\Phi$ distribution associated with HFAs (Fig.~\ref{fig:dphidist}) peaks at smaller values ($0^o-30^o$) when compared to the distribution of solar wind distribution rotation angles, which peaks at larger values \citep[$30^o-45^0$,][Fig.~2]{knetter04:_four_clust}. The $\gamma$ distribution associated with HFAs (Fig.~\ref{fig:cone}) shows a wide, empty cone around the Sun-Earth line, which is in contrast to the distributor of solar wind discontinuities, whose normals typically have small $\gamma$ angles \citep[Fig.~11]{knetter04:_four_clust}. We found only one normal vector within this cone in 2003 and a few others in 2006 and 2007. We observed this feature in the distribution of $\gamma$ (Fig.~\ref{fig:cone}). This finding strongly supports the earlier theoretical and simulation results that HFAs can only be formed if $43\degree\le\gamma\le83\degree$ \citep{lin02:_global, nemeth07:_partic_accel_at_inter_of, facsko08:_statis_study_of_hot_flow}. The distribution of $\Delta\Phi$ shows that HFAs can be formed if the magnetic field vector directional change is sufficiently large across the TD (Tab.~\ref{tab:tddataa}). Actually smaller values of $\Delta\Phi$ were also observed, which supports the theoretical results by \citep{lin02:_global, facsko08:_statis_study_of_hot_flow, facsko08:_clust_hot_flow_anomal_obser}. The distribution of TD normals for $\gamma > 45\degree$ is evenly distributed. We most often used ACE MAG measurements to determine TD normals in 2003, but had to mainly use Cluster FGM magnetic field data in 2007 because it was impossible to couple ACE and Cluster observations. The simulation was a better description of the events of 2006 than that of 2007. This turns out to be an advantage because the accuracy of $\gamma$ and $\Delta\Phi$ increased in 2006 and 2007.

\subsubsection{Estimations of HFA size}
\label{sec:size}

Cluster satellites cross HFAs but the time length of the event holds no information about the real size of the phenomena because the boundaries of the cavity rim are not in pressure balance \citep{thomsen86:_hot,lucek04:_clust} and the HFA also moves in the frame of the solar wind plasma. On the other hand, we have other valuable information: the time that the spacecraft spends inside the cavity gives a lower limit for the time of the existence of the HFA. One can calculate the error based on the measurements of four (or less) satellites. The size of the HFA must be estimated in another way. 

\begin{enumerate}
\item HFAs, hot diamagnetic cavities, are created by particle beams accelerated by the supercritical bow shock. The beam shares its energy through electromagnetic ion-ion beam instability. In fact, this beam creates Alfv\'en waves and these waves carry away a larger part of the energy; only 2/3 of the energy heats the plasma  \citep{thomas88:_evolut, thomas89:_three}. The propagation velocity of these waves does not exceed the Alfv\'en velocity so that twice the Alfv\'en speed multiplied by time of existence may give a rough estimate for the lower limit of the HFA size. \citet{schwartz85} determined the expansion speed of the cavity using ISEE-1 and ISEE-2 measurements, and the measured expansion speed was approximately the same as the estimated velocity. 

\item HFAs are formed by the interaction of the bow shock and a tangential discontinuity. In many numerical simulations \citep{burgess88:_collid_curren, lin02:_global, omidi07:_format} and observations \citep{lucek04:_clust} one can see that the HFA appears when the TD reaches the quasi-parallel region and remain while the TD sweeps the surface of the bow-show. We calculated the transit velocity of the tangential discontinuity on the surface of the bow shock using \citet{schwartz00:_condit}'s formula:
\begin{equation}
\label{eq:schwartz}
\mathbf{V}_{tr}=\frac{\mathbf{V}_{sw}\mathbf{n}_{cs}}{\sin^2\theta_{cs:bs}}\left(\mathbf{n}_{cs}-\cos\theta_{cs:bs}\mathbf{n}_{bs}\right), 
\end{equation}
where $\mathbf{V}_{tr}$ is the transient velocity, $\mathbf{V}_{sw}$ is the solar wind speed, $\mathbf{n}_{cs}$ is the normal of the tangential discontinuity (current sheet), $\mathbf{n}_{bs}$ is the normal of the bow shock, and $\theta_{cs:bs}$ is the angle between the two previously mentioned normals. The bow shock shape, position and normal were calculated by the model described in \citet{peredo95:_three_alfven_mach} as in the original paper which used ACE SWEPAM measurements. The solar wind vectors were determined by using Cluster CIS HIA measurements. This instrument operates only on Cluster SC1 and SC3. We obtained two estimates on the size of HFA. The obtained sizes are very similar after multiplying the velocity by the transition time of the spacecraft. 
\end{enumerate}
\begin{figure}[ht]
\centering
\epsfig{file=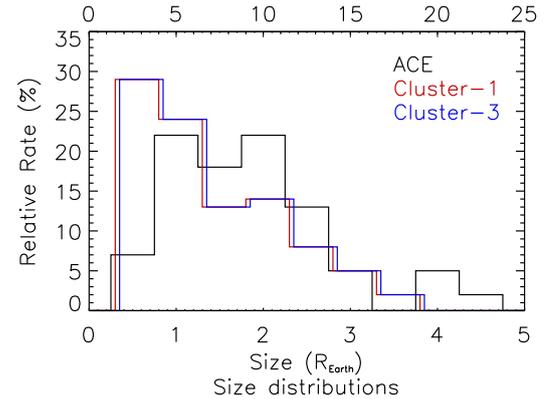,width=200pt}
\caption{The size distributions of HFAs estimated by Alfv\'en velocity and (solid line) the speed of the TD and bow shock intersection calculated by the solar wind measurements of Cluster-1 and -3 CIS HIA (red and blue line, scale drawn on top). The average sizes are $\left(1.9\pm1.0\right)\,R_{\mathrm{Earth}}$, $\left(7.0\pm4.3\right)\,R_{\mathrm{Earth}}$ and $\left(6.6\pm4.2\right)\,R_{\mathrm{Earth}}$, respectively.}
\label{fig:sizedistr}
\end{figure}
We estimated the size of HFAs and the errors based on the methods above. Each of them gives four results by four satellites. We took the average over the four points to be the size, with the standard deviation as the error. Unfortunately the CIS HIA aboard Cluster-1 and Cluster-3 provided unusually high temperatures close to the bow shock and so we used only the measurements of the ACE SWEPAM plasma instrument and the ACE MAG magnetometer to determine the properties of the plasma. For this reason only one size distribution from using the first method is given (Fig.~\ref{fig:sizedistr}). The average sizes and their errors are $\left(1.9\pm 1.0\right)\,R_{\mathrm{Earth}}$, $\left(7.0\pm 4.3\right)\,R_{\mathrm{Earth}}$ and $\left(6.6\pm4.2\right)\,R_{\mathrm{Earth}}$, respectively. The first result confirms the predictions of the \citeauthor{lin02:_global}'s theory however the second result seems to be much higher. Most of the distribution functions of the second estimation shows
  a value of approximately $5\,R_{\mathrm{Earth}}$. The reason for this higher average is the ``tail'' of the distribution at larger sizes. Unfortunately this size estimation is very sensitive to the errors of the different normals and velocity vectors (See: Eq.~\ref{eq:schwartz}) and often gives a very large size. After comparing the size distributions of two methods on Fig.~\ref{fig:sizedistr} one can see that most of their values do not differ by more then a factor of two. They are thus suitable for estimating the size of the phenomena. All side distributions are found to be very similar and the size-angle functions support the simulation results.

\subsubsection{Size-angle and size-speed scatter plots}
\label{sec:sizeangle}

Size-angle relations were reported in \citet{lin02:_global}. Furthermore we were informed about size-speed predictions \citep[personal communication]{lin07}.

\begin{figure*}[th]
\begin{center}
\epsfig{file=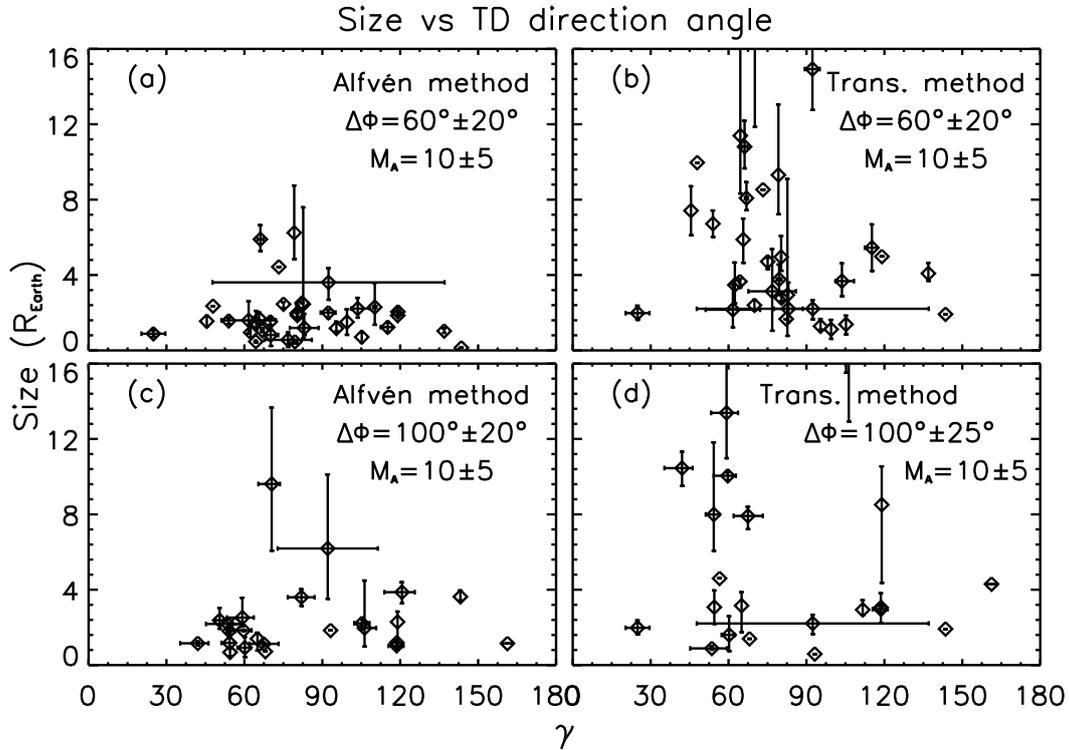,width=400pt} 
\end{center}
\caption{The size-$\gamma$ functions based on the size estimation by Alfv\'en Mach velocity on the left and the transition speed on the right. The fixed solar wind speed was shown in Alfv\'en Mach number. (a) $\Delta\Phi=60\degree\pm20\degree$ and $M_A=10\pm5$, (b) $\Delta\Phi=60\degree\pm20\degree$ and $M_A=10\pm5$, (c) $\Delta\Phi=100\degree\pm20\degree$ and $M_A=10\pm5$, (d) $\Delta\Phi=100\degree\pm25\degree$ and $M_A=10\pm5$. All Alfv\'en Mach numbers were calculated from the actual Alfv\'en velocity.}
\label{fig:sizeangleplots}
\end{figure*}

Fig.~\ref{fig:sizeangleplots} show the size-$\gamma$ correlations. The error of the size was calculated by the method described by Sec.~\ref{sec:size} and the error of the angles was estimated by the cross-product method: we calculated the direction for every single spacecraft, the average of these directions, and finally the error cone. The error of direction was not calculated where only one direction was obtained. It is very important to remark that the size depends not on one but three parameters. The size was plotted as a function of one parameter ($\gamma$) while the speed and $\Delta\Phi$ values were fixed. In fact, fixing a parameter means fixed angle intervals because these were real measurements and not theoretical models. We fixed the speed in Alf\'en-Mach number in the simulation as well. We chose these $\Delta\Phi$ intervals because these contains those points which were simulated by \citeauthor{lin02:_global} with $M_A=5$ and $\Delta\Phi=80\degree$. We obtained a maximum of the size-$\gamma$ scattered plot but not exactly at $\gamma=80^{o}$ in both cases as predicted \citep{lin02:_global}. The other panels also support the theory since a maximum is visible on every panel. When we plotted all points we obtained a ``cloud'' of points with a maximum value. 

\begin{figure*}[th]
\begin{center}
\epsfig{file=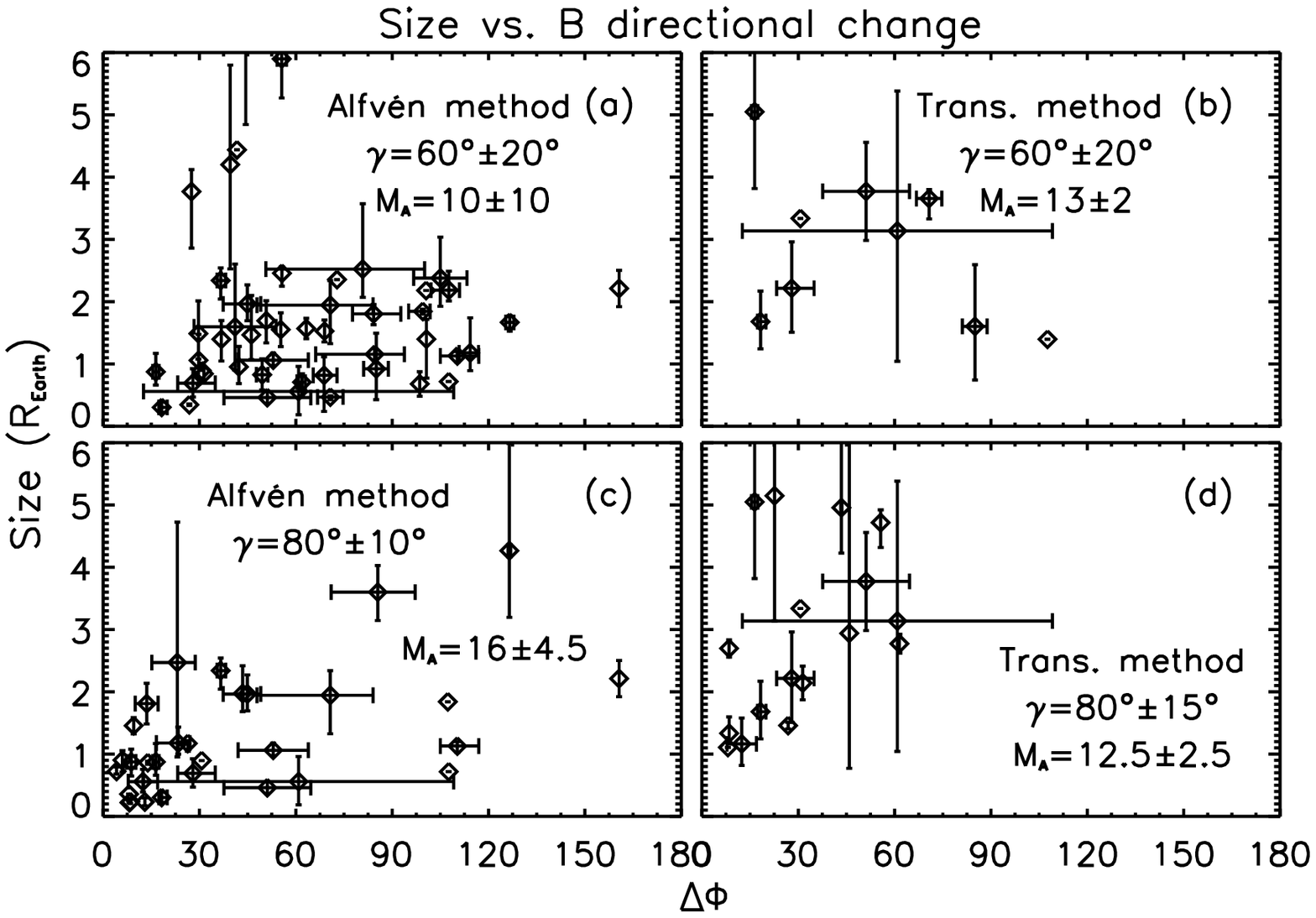,width=400pt} 
\end{center}
\caption{The size-$\Delta\Phi$ functions based on the size estimation by Alfv\'en Mach velocity on the left and the transition speed on the right. The fixed solar wind speed was shown in Alfv\'en Mach number. (a) $\gamma=60\degree\pm20\degree$ and $M_A=10\pm10$, (b) $\gamma=60\degree\pm20\degree$ and $M_A=13\pm2$, (c) $\gamma=80\degree\pm10\degree$ and $M_A=16\pm4.5$, (d) $\gamma=80\degree\pm15\degree$ and $M_A=12.5\pm2.5$. All Alfv\'en Mach numbers were calculated from the actual Alfv\'en velocity.} 
\label{fig:sizedirplots}
\end{figure*}

Fig.~\ref{fig:sizedirplots} presents the size-$\Delta\Phi$ functions where $\Delta\Phi$ is the change angle of magnetic field direction across the TD. The error of the size and angle were calculated the same way as at size-$\gamma$ functions. Here $\gamma$ and the solar wind speed were fixed and we used Alfv\'en Mach numbers. Here the bottom panels show the case studied in the simulation of \citet{lin02:_global}. All panels show monotonically increasing size-$\Delta\Phi$ functions, confirming simulation results. We obtain a set of points a dense region that increases to the larger sizes. 

\begin{figure}[b]
\centering
\epsfig{file=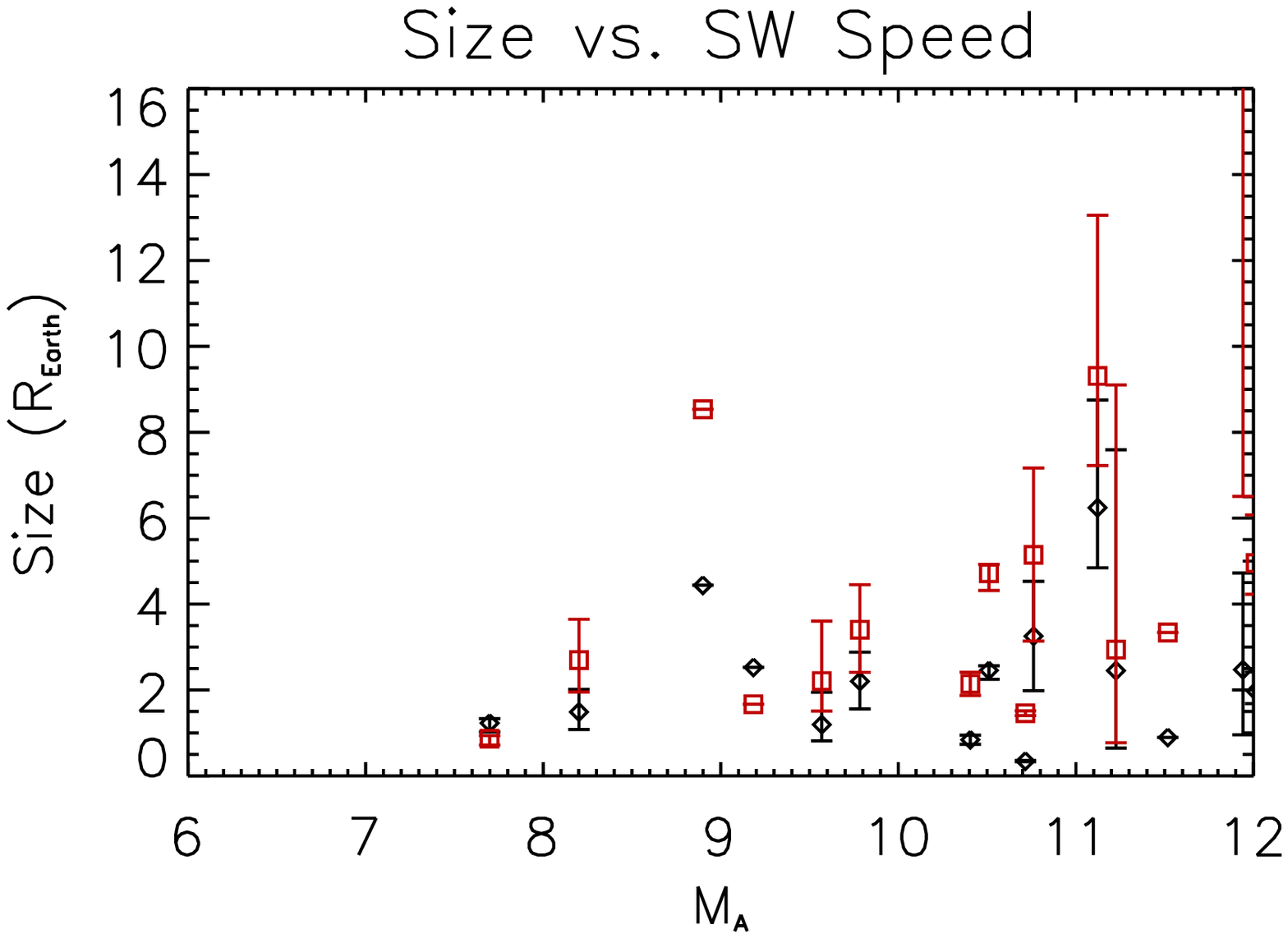,width=200pt} 
\caption{The size-velocity functions with Alfv\'en velocity calculated using ACE and crossing time measured by Cluster. The sizes were calculated using the method based on Alfv\'en speed (black) and the transition speed (red). The fixed solar wind speed was measured in units of Alfv\'en Mach number. $\gamma=80\degree\pm10\degree$ and $\Delta\Phi=40\degree\pm20\degree$. All Alfv\'en Mach numbers were calculated from the actual Alfv\'en velocity.}
\label{fig:sizespeedplots}
\end{figure}

In Fig.~\ref{fig:sizespeedplots} the dependence of HFA size on velocity is visible in several fixed angle intervals. Solar wind speed was measured in Alfv\'en Mach number value. The size was estimated based on the Alfv\'en speed method (black) and by calculating the velocity of the intersection line of the TD and the bow shock (red). The angular dependence of size was studied in a fixed intervals around $\gamma=80\degree$ and $\Delta\Phi=40\degree$ angles and the size is the monotonically growing function of the Alfv\'en Mach number. 

\subsection{Speed distributions}
\label{sec:speeddistr}

\begin{figure*}[th]
\centering
\epsfig{file=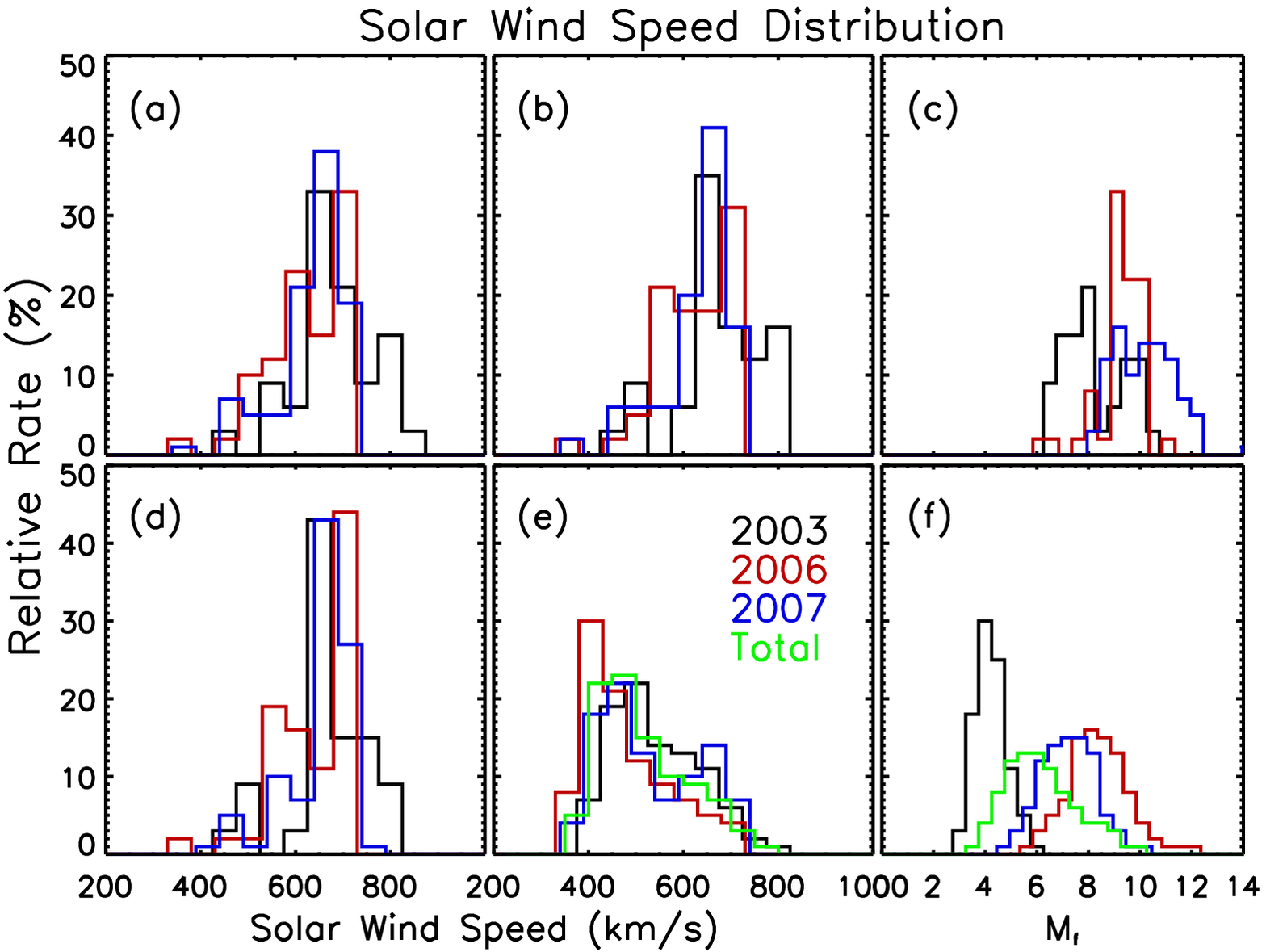,width=400pt}
\caption{Solar wind speed distribution measured by Cluster and ACE spacecraft. Black, red, blue and green refers to measurements in 2003, 2006, 2007 and 1998-2008, respectively. The figure shows the solar wind speed distribution measured by (a) Cluster-1 CIS HIA during HFA formation. (b) by Cluster-3 CIS HIA, and (d) by ACE SWEPAM; it also shows. Fast magnetosonic Mach number distribution calculated using ACE MAG and SWEPAM data during HFA formation (c), solar wind speed distribution measured by ACE SWEPAM from February to April, 2003, December 2005-April, 2006 and January-April, 2007 and 1998-2008 (e), and fast-magnetosonic Mach-number distribution (f).}
\label{fig:vSWdistr}
\end{figure*}

\begin{table*}[bht]
\centering
\begin{tabular}[c]{r||r@{$\pm$}l|r@{$\pm$}l|r@{$\pm$}l|c}
solar wind speed $\left(km/s\right)$ & \multicolumn{2}{c|}{2003} & \multicolumn{2}{c|}{2006} & \multicolumn{2}{c|}{2007} & Fig \\
\hline
during HFA formation by C1 & 680 & 86 & 614 & 84 & 613 & 80 & \ref{fig:vSWdistr}a \\
 by C3                     & 671 & 92 & 614 & 82 & 613 & 78 & \ref{fig:vSWdistr}b \\
 by ACE                    & 666 & 84 & 626 & 85 & 634 & 71 & \ref{fig:vSWdistr}d \\
$M_f$ numbers by ACE       & 8.2 & 1.2 & 9.1 & 1.0 & 9.9 & 1.1 & \ref{fig:vSWdistr}c \\
\hline
in 3/4 months period by ACE & 546 & 97 & 477 & 97 & 512 & 102 & \ref{fig:vSWdistr}e \\
\hline
between 1998-2003/2008 by ACE & 492 & 102 & \multicolumn{4}{c|}{498$\pm$101} & \ref{fig:vSWdistr}e \\
$M_f$ numbers by ACE          & 5.5 & 1.4 & \multicolumn{4}{c|}{6.2$\pm$1.7} & \ref{fig:vSWdistr}f \\
\hline \hline
$\Delta M_f$               & \multicolumn{2}{c|}{2.7} & \multicolumn{2}{c|}{2.9} & \multicolumn{2}{c|}{3.7} & \\
\end{tabular}
\caption{Solar wind speed, fast magnetosonic Mach number mean values, and their deviations measured by Cluster CIS and ACE SWEPAM. The last column gives the figure numbers shown on Fig.~\ref{fig:vSWdistr}.}
\label{tab:sw}
\end{table*}

We observed in our previous work \citep{kecskemety06:_distr_rapid_clust} that the value of the solar wind speed is close to the average $\sim$400\,km/s but it is higher before HFAs are observed ($\sim$600\,km/s). We have studied this point in more detail here. The speed distributions were calculated here we used Cluster SC1 and SC3 CIS HIA; complemented by ACE SWEPAM data measured in longer time intervals to obtain better statistics.  We recorded these solar wind speed values again when we used 5-10 minute or even 30 minute long intervals before the bow shock. We calculated the average, its scatter and plotted the distribution (Tab.~\ref{tab:sw}, Fig.~\ref{fig:vSWdistr}). We determined the time when the TD (which caused the HFA) crossed the position of ACE satellite and we determined the average solar wind parameters from ACE SWEPAM measurements. These results are in good agreement with earlier Cluster observations \citep{facsko08:_statis_study_of_hot_flow, facsko08:_clust_hot_flow_anomal_obser}. 

\begin{figure*}[th]
\centering
\begin{tabular}[t]{rl}
\multicolumn{2}{c}{\epsfig{file=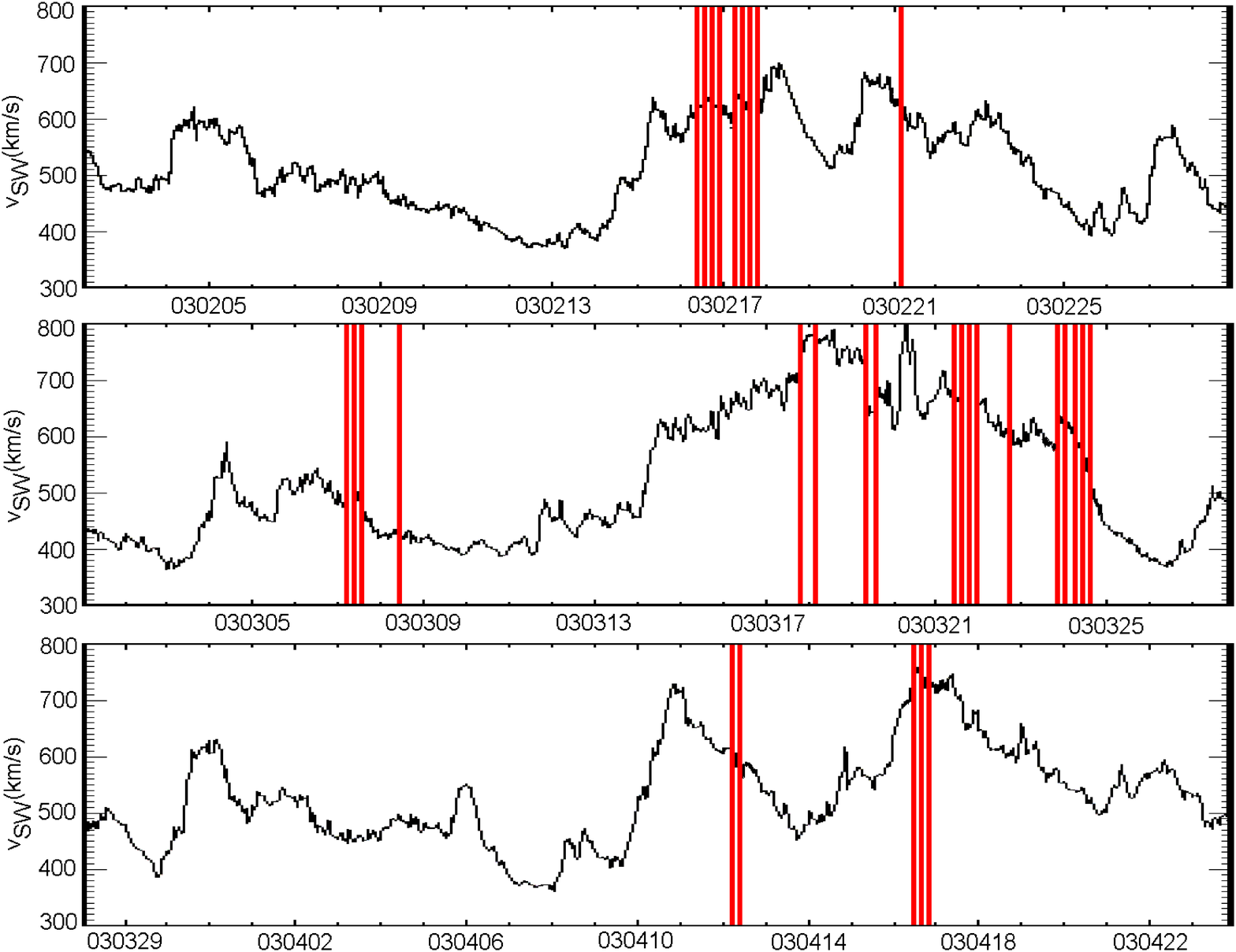,width=300pt}} \\
\epsfig{file=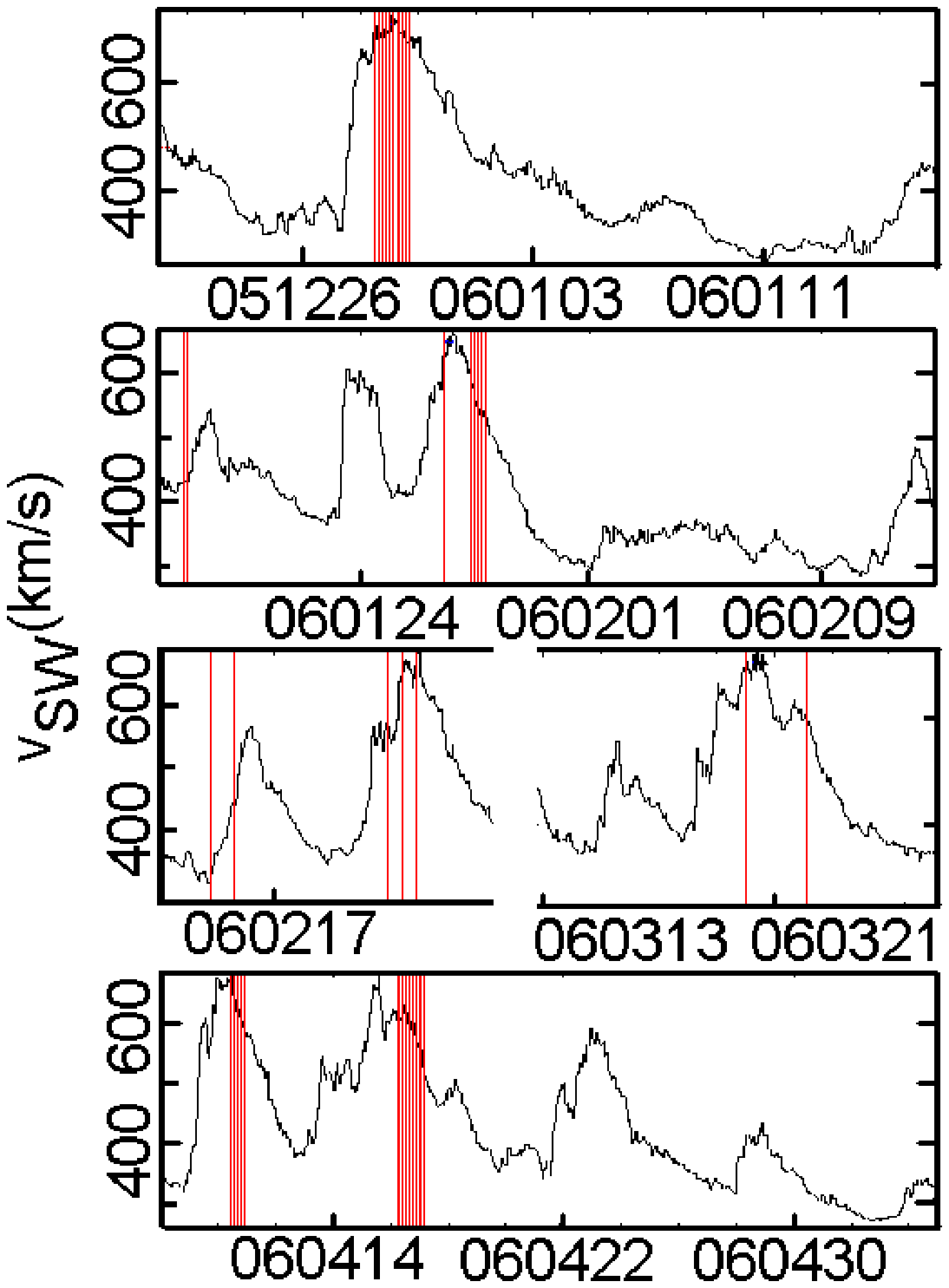,width=200pt} &
\epsfig{file=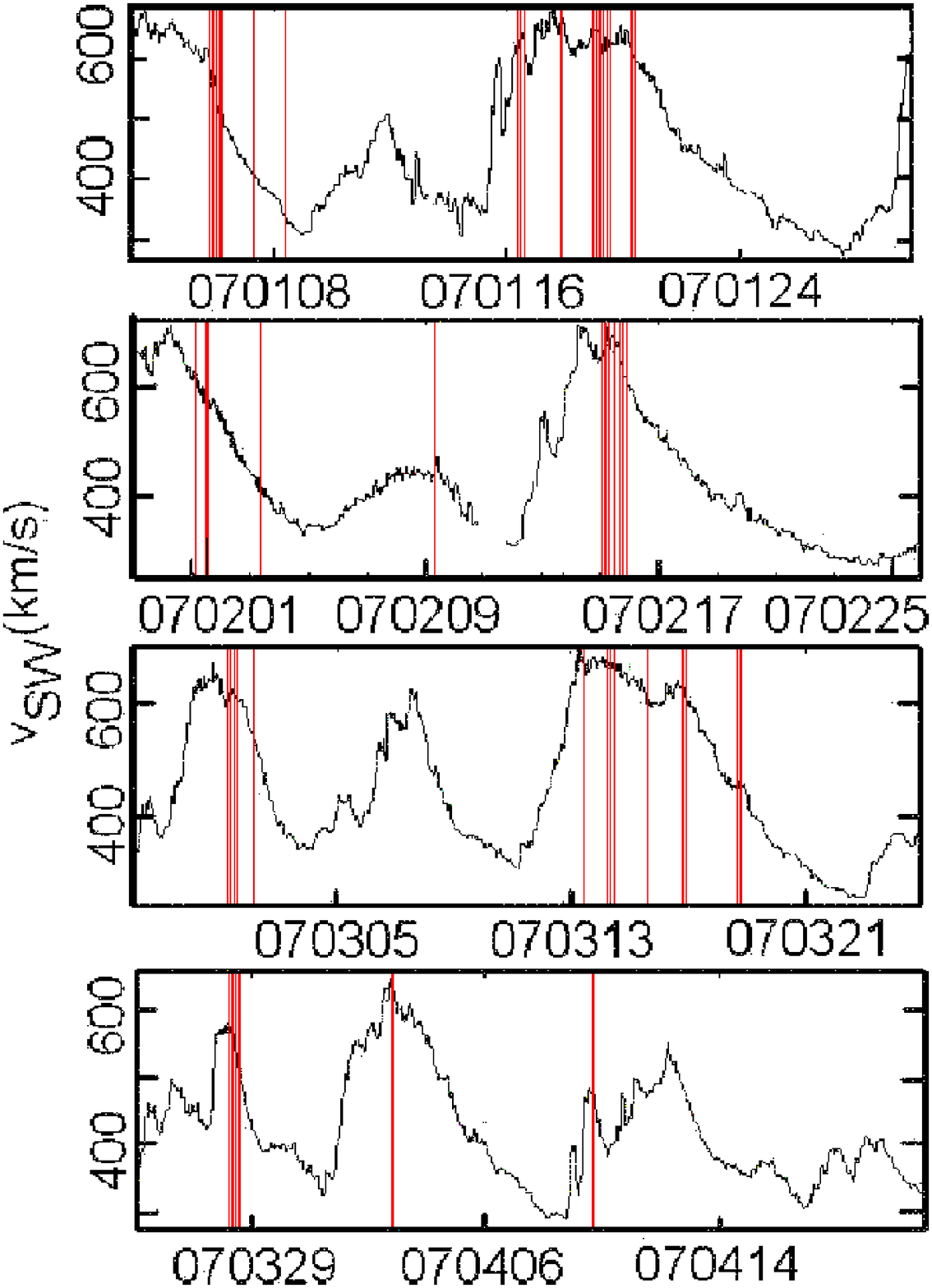,width=200pt} \\
\end{tabular}
\caption{1 hour averaged solar wind speed; the vertical red lines give the time of HFAs. The top, bottom left and right figures were measured by ACE SWEPAM instrument in 2003, 2006 and 2007, respectively. The connection between the fast solar wind regions and the HFAs is evident.}
\label{fig:vSWvel}
\end{figure*}

These speeds are obviously higher than the long-term averaged solar wind speed (Fig.~\ref{fig:vSWdistr}a, b, d), and a peak appears on the distribution between 400\,km/s and 800\,km/s measured instead of the expected 400\,km/s or 800\,km/s peaks measured by Ulysses \citep{mccomas03}, but it is in question whether this difference is really significant. The average speed for the full-studied time period using ACE SWEPAM (Fig.~\ref{fig:vSWdistr}e, black line) was $\left(546 \pm 97\right)\,km/s$ in 2003. Actually, the solar wind speed was higher throughout the studied period in 2003 (Fig.~\ref{fig:vSWvel}). Measurements of ACE from 1998 to 2008 (Fig.~\ref{fig:vSWdistr}e, green line) yielded $\left(498 \pm 101\right)\,km/s$ suggesting that \textit{during HFA formation the typical solar wind speed is higher than the average value by almost 200\,km/s than the average value}. It seems that \textit{the presence of a fast solar wind is a necessary condition of the formation of HFAs}. This is obvious when one looks at the bottom panel of Fig.~\ref{fig:vSWvel}, where we plotted the studied interval using 1\,hour averaged solar wind speed. The HFAs marked by vertical lines and their positions all appear in fast solar wind regimes. In fact almost all HFA events appeared in the same co-rotating region \citep{facsko08:_statis_study_of_hot_flow, facsko08:_clust_hot_flow_anomal_obser}. The frequency of fast solar wind beams in the Ecliptic depends on the solar cycle. The frequency of HFAs is thus expected to depend on solar cycle. After processing the measurements in 2006 and 2007 this cannot be confirmed because the average number of HFAs is about 2 HFAs/day with large scatter ($2.2\pm1.2$, $2.5\pm1.4$ and $2.1\pm1.5$ in 2003, 2006 and 2007, respectively) so there is no significant difference during the different seasons. There were several longer HFA series in 2006 and 2007 but not in is 2003. The difference between the solar wind speeds were high -- \textit{~130\,km/s} -- but not as high as in 2003. Based on three years of measurements of we can conclude that the higher solar wind speed might be an important requirement for the HFA formation mechanism. We found only a few HFAs out of the fast solar wind co-rotating regions. 

Fig.~\ref{fig:vSWdistr}c shows a more unexpected result. The figure shows the distribution of the fast-magnetosonic Mach numbers during HFA formation. The Mach numbers are very high, with $M_{f}\ge6$ in 2003, this can also be observed in 2006 and 2007 where the difference between them is even greater. This is made more obvious if we compare this distribution to the distribution calculated by ACE SWEPAM and MAG measurements for the studied interval and all measurements of ACE (Fig.~\ref{fig:vSWdistr}f). Both longer periods show that these high Mach numbers are very rare \citep{facsko08:_statis_study_of_hot_flow}. The HFAs are not only Earth-specific features \citep{oeieroset01:_hot_martian}. The Mach numbers are in general much larger in the outer Solar System, since the propagation speed of fast magnetosonic waves is lower due to the weaker magnetic field. This fact suggests that HFA events might be even more frequent at Saturn, for instance the other giant planets in the Solar System.

\subsection{Solar wind density and pressure}
\label{sec:pressure}

\begin{figure*}[t]
\centering
\epsfig{file=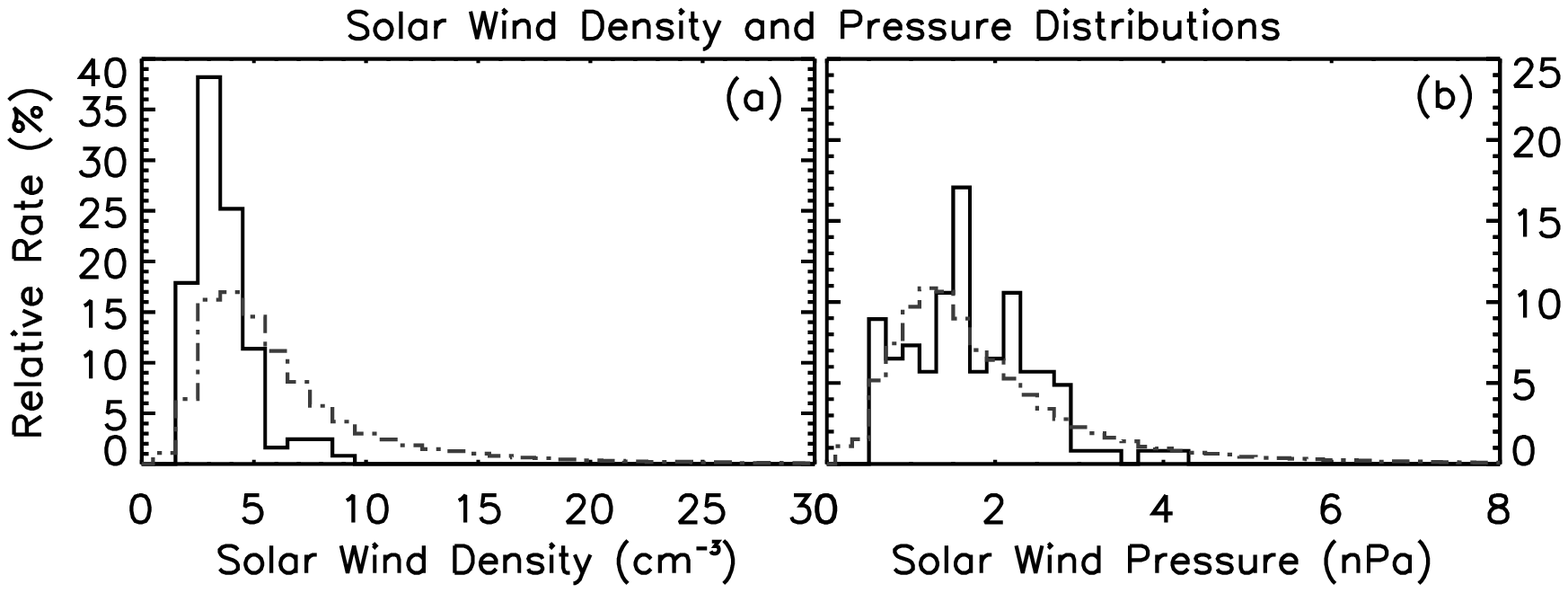,width=400pt}
\caption{(a) Solar wind particle density distribution during HFA events (dash-dotted line) using ACE SWEPAM measurements from 1998 to 2008 (solid line). (b) Solar wind pressure in the same time intervals.}
\label{fig:vSWp}
\end{figure*}

Several HFA events are shown on Fig.~\ref{fig:vSWvel} when the solar wind velocity is above average, but which do not have very large values. The higher solar wind velocity seems to be a necessary condition of forming HFAs so these exceptions look strange. We studied parameters, one of which was solar wind particle density.  (Fig.~\ref{fig:vSWp}a). We noticed that the particle density is below the average during an HFA formation, at $3.6\pm1.4\,cm^{-3}$ instead of the long-term average value of $6.9\pm4.2\,cm^{-3}$ (based on the ACE SWEPAM 1\,hour average data series measured between 1998 and 2008). This observation is not surprising since the solar wind pressure is approximately constant. Thus, if the solar wind velocity is higher, the density is expected to be lower.

The other studied parameter was the solar wind pressure. We also calculated distribution function, which suggested lower pressure during HFA formation than the average of all measurements of ACE from 1998 to 2008. It was $1.7\pm 0.8\,nP$ instead of the $1.9\pm 1.2\,nPa$ (Fig.~\ref{fig:vSWp}b). In our opinion this difference is not significant. Unfortunately the high solar wind pressure does not seem to be a condition of HFA formation in the case of those few events when the solar wind speed is not too large. 

\subsection{Schwartz et. al.'s condition}
\label{sec:schwartz}

\begin{figure}[hb]
\centering
\epsfig{file=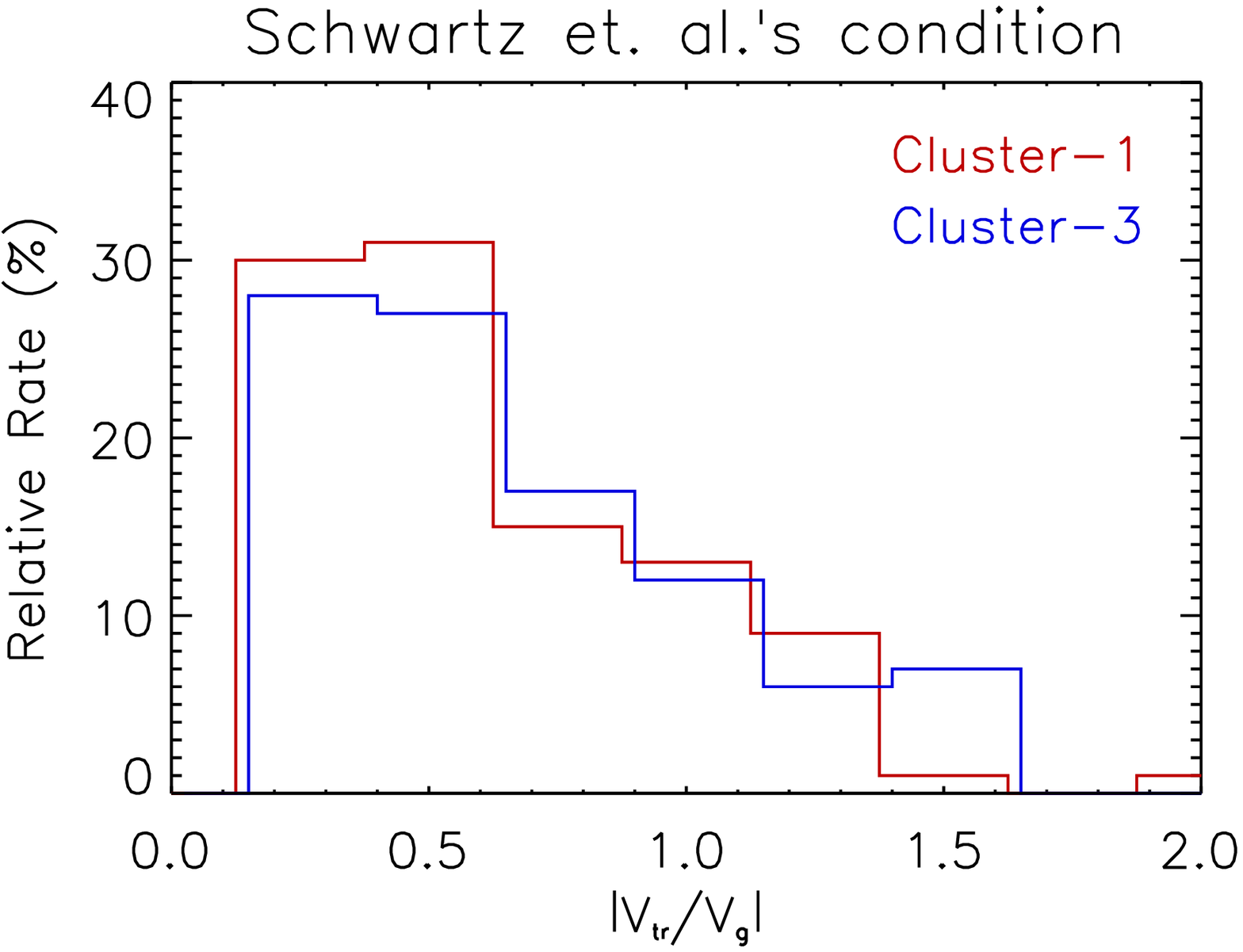,width=200pt}
\caption{The distribution of the rate given by Eq.~\ref{eq:schwartzc}. We use both Cluster SC1 and SC3 CIS HIA measurements to determine the necessary vectors in the formula. The red and blue lines show the distribution based on Cluster-1 and -3 measurements.}
\label{fig:schwartz}
\end{figure}

We have checked whether the \citet{schwartz00:_condit} condition is valid for our HFA events. The above discussed analysis of HFA events in the spring 2003, 2006 and 2007 seasons confirmed and extended our earlier results based on the study of HFA events in spring 2003. These showed that higher solar wind speed is an important condition of HFA formation. This feature restricts the formula of \citet{schwartz00:_condit} because the TD must slowly sweep the bow shock which is possible for only a very limited geometrical condition. Besides of these limitations our events also confirm the following results:
\begin{equation}
\label{eq:schwartzc}
\left|\frac{V_{tr}}{V_g}\right|=\frac{\cos\theta_{cs:sw}}{2\cos\theta_{bs:sw}\sin\theta_{B_n}\sin\theta_{cs:bs}}<1, 
\end{equation}
where $V_{tr}$ is the transit velocity of the current sheet along the bow shock, $V_g$ is the gyration speed, $\theta_{cs:sw}$, $\theta_{bs:sw}$ and $\theta_{cs:bs}$ are the angles between the discontinuity normal, solar wind velocity and the bow shock, and finally $\theta_{B_n}$ is the angle between the magnetic field and bow shock normal. The necessary vectors were calculated using Cluster SC1 and SC3 CIS HIA measurements (Fig.~\ref{fig:schwartz}). We found that the transition speed is most often as low as expected by the formula of \citet{schwartz00:_condit}. This formula usually gives a value of less than 1 one during HFA formation. Here the formula often gives a greater value than one; however, this study also confirms that HFA formation also depends on the geometry of the shock, the discontinuity, and the solar wind velocity. 

\section{Discussion}
\label{sec:discus}

Our resulting value of size estimation, the shape of size-angle and size-velocity distributions, as well as the function of $\Delta\Phi$ and $\gamma$, confirm previous predictions of numerical simulations. The large number of events, as well as the higher solar wind speed and Mach number are new results although \citet{koval05} had made similar observations using INTERBALL-1 and MAGION-4 spacecraft. (That study was performed using magnetosheath observation instead of upstream measurements.) All our observations agree well with current theories and simulations. 

The high solar wind velocity as an essential condition is logical and acceptable because particles of the beam which form the HFA are accelerated at the supercritical bow shock. Here, the particles are forced to return to the foreshock region approximately with solar wind speed, but antiparalel to solar wind velocity \citep{gosling85:_ion, kennel85, scholer93:_two, tanaka83, quest:_hybrid_simul}. This process causes the heating of the region and the energy dissipation of the flow, and forms the beam which creates the HFA. The higher the speed of the solar wind, the higher the energy of the reflected beams. Moreover, analytical calculations by \citet{nemeth07:_partic_accel_at_inter_of} (which study the possible particle trajectories of trapped ions in the vicinity of shock-discontinuity crossings) suggest high solar wind speed as a favorable condition of particle reflection. Unfortunately no numerical simulation thus for can predict this condition, probably because these simulations are constrained into 2 spatial dimensions. 3D hybrid simulations may be able to predict the high solar wind speed condition.

The $\gamma$ distribution and the size maxima of size-$\gamma$ functions \citep{lucek04:_clust, schwartz00:_condit} are explained as follows: acceleration needs time and the TD must approach the bow shock. If the angle is large then it approaches slower and there is more time for acceleration. Beyond at given angle particles do not bounce back and nothing forms. The situation is different in the case of growing size-$\Delta\Phi$ functions. \citet{lin07} suggests that the electric field depends on this angle, so larger $\Delta\Phi$ generates larger electric field which focuses particles to the TD. It is well known that the acceleration happens between the TD and the quasi-parallel shock. When the TD reaches the quasi-parallel region of the bow shock or when the TD changes the magnetic field direction, the particles -- which form the beam -- can escape from the trap, which gives rise to the phenomenon. Larger $\Delta\Phi$ causes longer acceleration time, which can explain the growing size-$\Delta\Phi$ functions. 

The reason of the growing size-speed function can be the following: the beam that creates the HFAs is accelerated at the supercritical bow shock. This result is not surprising because their acceleration depends on the bow shock structure. A small amount of particles turns back and enters the region in front of the bow shock, the foreshock region or the region between the bow shock and the TD. TD occurs when the HFA is formed. The higher the velocity of the solar wind, the higher the speed of particles and size of the phenomenon. This trend can be seen on the Fig.~\ref{fig:sizespeedplots}, however it is not very obvious. 

\conclusions[Summary and conclusions]
\label{sec:conc}

Earlier we showed that HFAs are not as rare a phenomenon as it was a thought prior to Cluster \citep{kecskemety06:_distr_rapid_clust}. If a TD appears and the spacecraft are in the right position then the event can be observed with high probability if several special conditions are fulfilled. The numerous new HFA observations also confirm this opinion. 
\begin{enumerate}
\item The most important condition is the larger solar wind velocity, which is typically much higher than the average speed. The differences were approximately 160\,km/s in 2003, and approximately 130\,km/s in 2006 and 2007. 
\item The high fast magnetosonic Mach number is also a preferable condition for HFA formation. No events were found below $M_f=6$ in 2003, and this limit increased in 2006 and 2007. 
\item The pressure is irrelevant with respect to HFA formation. The solar wind particle density before the HFA events is lower than the average value of the solar wind density. 
\item The angle between the TD normal ($\gamma$) and Earth-Sun direction must be greater than $45^{o}$. Very few events were observed with $\gamma<45^o$. 
\item The directional change of magnetic field within the TD ($\Delta\Phi$) must be large. The average value was approximately $70\degree$ based on 124 events. 
\item Our size estimations do not contradict previous simulation results. We estimated $2-3\,R_{\mathrm{Earth}}$ size using one method; the other method gave larger sizes in the range of $1\,R_{\mathrm{Earth}}$. The differences can be explained with the high sensitivity of the methods to the accuracy of the measurements. 
\item The size-angle and size-speed plots of \citet{lin02:_global} were reproduced in good agreement with the predictions. 
\item The conditions were mostly quasi-parallel during HFA formation, which is unexpected because the HFA determination decreases the number of quasi-parallel cases. So our HFA observations confirm the previous simulation result of \citet{omidi07:_format} and showed that HFAs appear where the quasi-perpendicular condition turns to quasi-parallel. Furthermore, the particles of the beam escape in the quasi-parallel part of the bow shock.  
\item We also confirmed the suggestion of \citet{schwartz00:_condit}, namely that the transition velocity of the HFA at the bow shock must be slow. Furthermore, our new result does not contradict to the formula presented in that paper (Eq.~\ref{eq:schwartzc}).
\end{enumerate}
We have determined the typical size of HFAs in two different ways. The number of HFAs does not depend on solar activity, only on the time of periods when the solar wind velocity is high. We compared within the theoretical predictions and proved that they are correct in 2003, 2006 and 2007 when the Cluster fleet separation was large. All observations agree well with current theories and confirm the simulation results. We also publish here the detected events and their parameters. We hope they will be used to further studies, for example, THEMIS-Cluster multi-multispacecraft observations or further statistical investigations beyond and inside the bow shock. 

The reason why the high solar wind velocity is necessary for HFA formation was not explained in detail. Further -- probably 3D hybrid -- simulations are necessary to clarify the theoretical background of this behavior. 

\begin{acknowledgements}
The authors thank the ACE MAG and SWEPAM working teams for the magnetic field and plasma data; furthermore the authors are also very grateful to Mariella T\'atraallyay for providing high resolution Cluster FGM data files. The present work was supported by the OTKA grant K75640 of the Hungarian Scientific Research Fund. G\'abor Facsk\'o thanks Pierrette Decreau and Robert Ferdman for their help in improving the English of this paper.
\end{acknowledgements}

 \bibliographystyle{copernicus}
\bibliography{angeo-2008-0009-tx}

\begin{thebibliography}{37}
\providecommand{\natexlab}[1]{#1}
\providecommand{\url}[1]{{\tt #1}}
\providecommand{\urlprefix}{URL }
\expandafter\ifx\csname urlstyle\endcsname\relax
  \providecommand{\doi}[1]{doi:\discretionary{}{}{}#1}\else
  \providecommand{\doi}{doi:\discretionary{}{}{}\begingroup
  \urlstyle{rm}\Url}\fi

\bibitem[{{Balogh} et~al.(2001){Balogh}, {Carr}, {Acu{\~n}a}, {Dunlop}, {Beek},
  {Brown}, {Forna{\c c}on}, {Georgescu}, {Glassmeier}, {Harris}, {Musmann},
  {Oddy}, and {Schwingenschuh}}]{balogh01:_clust_magnet_field_inves}
{Balogh}, A., {Carr}, C.~M., {Acu{\~n}a}, M.~H., {Dunlop}, M.~W., {Beek},
  T.~J., {Brown}, P., {Forna{\c c}on}, K.-H., {Georgescu}, E., {Glassmeier},
  K.-H., {Harris}, J., {Musmann}, G., {Oddy}, T., and {Schwingenschuh}, K.:
  {The Cluster Magnetic Field Investigation: overview of in-flight performance
  and initial results}, \anngeo, 19, 1207--1217, 2001.

\bibitem[{{Burgess} and {Schwartz}(1988)}]{burgess88:_collid_curren}
{Burgess}, D. and {Schwartz}, S.~J.: {Colliding plasma structures - Current
  sheet and perpendicular shock}, \jgr, 93, 11\,327--11\,340, 1988.

\bibitem[{{CIS Team}(1997-present)}]{caveat_for_data_suppl_by}
{CIS Team}: CAVEATS for the Data supplied by the CIS Experiment Onboard the
  Cluster Spacecraft, web page, 1997-present.

\bibitem[{{Facsk\'o} et~al.(){Facsk\'o}, {T\'atrallyay}, {Erd\H{o}s}, and
  {Dandouras}}]{facsko08:_clust_hot_flow_anomal_obser}
{Facsk\'o}, G., {T\'atrallyay}, M., {Erd\H{o}s}, G., and {Dandouras}, I.:
  Cluster hot flow anomaly observations during solar cycle minimum, in:
  Proceedings of the 15th Cluster Workshop \& Cluster Active Archive School,
  Springer Verlag.

\bibitem[{{Facsk{\'o}} et~al.(2008){Facsk{\'o}}, {Kecskem{\'e}ty}, {Erd{\H
  o}s}, {T{\'a}trallyay}, {Daly}, and
  {Dandouras}}]{facsko08:_statis_study_of_hot_flow}
{Facsk{\'o}}, G., {Kecskem{\'e}ty}, K., {Erd{\H o}s}, G., {T{\'a}trallyay}, M.,
  {Daly}, P.~W., and {Dandouras}, I.: {A statistical study of hot flow
  anomalies using Cluster data}, Advances in Space Research, 41, 1286--1291,
  \doi{10.1016/j.asr.2008.02.005}, 2008.

\bibitem[{{Gosling} and {Robson}(1985)}]{gosling85:_ion}
{Gosling}, J.~T. and {Robson}, A.~E.: {Ion reflection, gyration, and
  dissipation at supercritical shocks}, Washington DC American Geophysical
  Union Geophysical Monograph Series, 35, 141--152, 1985.

\bibitem[{{Kecskem{\'e}ty} et~al.(2006){Kecskem{\'e}ty}, {Erd{\H o}s},
  {Facsk{\'o}}, {T{\'a}trallyay}, {Dandouras}, {Daly}, and
  {Kudela}}]{kecskemety06:_distr_rapid_clust}
{Kecskem{\'e}ty}, K., {Erd{\H o}s}, G., {Facsk{\'o}}, G., {T{\'a}trallyay}, M.,
  {Dandouras}, I., {Daly}, P., and {Kudela}, K.: {Distributions of suprathermal
  ions near hot flow anomalies observed by RAPID aboard Cluster}, \aisr, 38,
  1587--1594, \doi{10.1016/j.asr.2005.09.027}, 2006.

\bibitem[{{Kennel} et~al.(1985){Kennel}, {Edmiston}, and {Hada}}]{kennel85}
{Kennel}, C.~F., {Edmiston}, J.~P., and {Hada}, T.: {A quarter century of
  collisionless shock research}, Washington DC American Geophysical Union
  Geophysical Monograph Series, 34, 1--36, 1985.

\bibitem[{{Knetter} et~al.(2004){Knetter}, {Neubauer}, {Horbury}, and
  {Balogh}}]{knetter04:_four_clust}
{Knetter}, T., {Neubauer}, F.~M., {Horbury}, T., and {Balogh}, A.: {Four-point
  discontinuity observations using Cluster magnetic field data: A statistical
  survey}, \jgr, 109, 6102, \doi{10.1029/2003JA010099}, 2004.

\bibitem[{{Koval} et~al.(2005){Koval}, {{\v S}afr{\'a}nkov{\'a}}, and {N{\v
  e}me{\v c}ek}}]{koval05}
{Koval}, A., {{\v S}afr{\'a}nkov{\'a}}, J., and {N{\v e}me{\v c}ek}, Z.: {A
  study of particle flows in hot flow anomalies}, \planss, 53, 41--52,
  \doi{10.1016/j.pss.2004.09.027}, 2005.

\bibitem[{{Lin}(2002)}]{lin02:_global}
{Lin}, Y.: {Global hybrid simulation of hot flow anomalies near the bow shock
  and in the magnetosheath}, \planss, 50, 577--591, 2002.

\bibitem[{Lin(2007)}]{lin07}
Lin, Y.: personal communication, 2007.

\bibitem[{{Lucek} et~al.(2004){Lucek}, {Horbury}, {Balogh}, {Dandouras}, and
  {R{\`e}me}}]{lucek04:_clust}
{Lucek}, E.~A., {Horbury}, T.~S., {Balogh}, A., {Dandouras}, I., and
  {R{\`e}me}, H.: {Cluster observations of hot flow anomalies}, \jgr, 109,
  6207, \doi{10.1029/2003JA010016}, 2004.

\bibitem[{{McComas} et~al.(1998){McComas}, {Bame}, {Barker}, {Feldman},
  {Phillips}, {Riley}, and
  {Griffee}}]{mccomas98:_solar_wind_elect_proton_alpha}
{McComas}, D.~J., {Bame}, S.~J., {Barker}, P., {Feldman}, W.~C., {Phillips},
  J.~L., {Riley}, P., and {Griffee}, J.~W.: {Solar Wind Electron Proton Alpha
  Monitor (SWEPAM) for the Advanced Composition Explorer}, Space Science
  Reviews, 86, 563--612, \doi{10.1023/A:1005040232597}, 1998.

\bibitem[{{McComas} et~al.(2003){McComas}, {Elliott}, {Schwadron}, {Gosling},
  {Skoug}, and {Goldstein}}]{mccomas03}
{McComas}, D.~J., {Elliott}, H.~A., {Schwadron}, N.~A., {Gosling}, J.~T.,
  {Skoug}, R.~M., and {Goldstein}, B.~E.: {The three-dimensional solar wind
  around solar maximum}, \grl, 30, 24--1, \doi{10.1029/2003GL017136}, 2003.

\bibitem[{{N\'emeth}(2007)}]{nemeth07:_partic_accel_at_inter_of}
{N\'emeth}, Z.: Particle acceleration at the interaction of shocks and
  discontinuities, in: Proceedings of the 30th International Cosmic Ray
  Conference, 2007.

\bibitem[{{{\O}ieroset} et~al.(2001){{\O}ieroset}, {Mitchell}, {Phan}, {Lin},
  and {Acu{\~n}a}}]{oeieroset01:_hot_martian}
{{\O}ieroset}, M., {Mitchell}, D.~L., {Phan}, T.~D., {Lin}, R.~P., and
  {Acu{\~n}a}, M.~H.: {Hot diamagnetic cavities upstream of the Martian bow
  shock}, \grl, 28, 887--890, \doi{10.1029/2000GL012289}, 2001.

\bibitem[{{Omidi} and {Sibeck}(2007)}]{omidi07:_format}
{Omidi}, N. and {Sibeck}, D.~G.: {Formation of hot flow anomalies and solitary
  shocks}, Journal of Geophysical Research (Space Physics), 112, 1203,
  \doi{10.1029/2006JA011663}, 2007.

\bibitem[{{Onsager} et~al.(1991){Onsager}, {Winske}, and
  {Thomsen}}]{onsager91:_inter_reflec}
{Onsager}, T.~G., {Winske}, D., and {Thomsen}, M.~F.: {Interaction of a
  finite-length ion beam with a background plasma - Reflected ions at the
  quasi-parallel bow shock}, \jgr, 96, 1775--1788, 1991.

\bibitem[{{Peredo} et~al.(1995){Peredo}, {Slavin}, {Mazur}, and
  {Curtis}}]{peredo95:_three_alfven_mach}
{Peredo}, M., {Slavin}, J.~A., {Mazur}, E., and {Curtis}, S.~A.:
  {Three-dimensional position and shape of the bow shock and their variation
  with Alfvenic, sonic and magnetosonic Mach numbers and interplanetary
  magnetic field orientation}, \jgr, 100, 7907--7916, 1995.

\bibitem[{{Quest}(1989)}]{quest:_hybrid_simul}
{Quest}, K.: Hybrid Simulation, in: Tutorial Courses: Third International
  School for Space Simulation, Toulouse, France, edited by {Lembege}, B.,
  {Eastwood}, J., and {Nepadues}, E., p. 177, 1989.

\bibitem[{{R{\`e}me} et~al.(2001){R{\`e}me}, {Aoustin}, {Bosqued}, {Dandouras},
  {Lavraud}, {Sauvaud}, {Barthe}, {Bouyssou}, {Camus}, {Coeur-Joly}, {Cros},
  {Cuvilo}, {Ducay}, {Garbarowitz}, {Medale}, {Penou}, {Perrier}, {Romefort},
  {Rouzaud}, {Vallat}, {Alcayd{\'e}}, {Jacquey}, {Mazelle}, {D'Uston},
  {M{\"o}bius}, {Kistler}, {Crocker}, {Granoff}, {Mouikis}, {Popecki},
  {Vosbury}, {Klecker}, {Hovestadt}, {Kucharek}, {Kuenneth}, {Paschmann},
  {Scholer}, {Sckopke}, {Seidenschwang}, {Carlson}, {Curtis}, {Ingraham},
  {Lin}, {McFadden}, {Parks}, {Phan}, {Formisano}, {Amata},
  {Bavassano-Cattaneo}, {Baldetti}, {Bruno}, {Chionchio}, {di Lellis},
  {Marcucci}, {Pallocchia}, {Korth}, {Daly}, {Graeve}, {Rosenbauer},
  {Vasyliunas}, {McCarthy}, {Wilber}, {Eliasson}, {Lundin}, {Olsen}, {Shelley},
  {Fuselier}, {Ghielmetti}, {Lennartsson}, {Escoubet}, {Balsiger}, {Friedel},
  {Cao}, {Kovrazhkin}, {Papamastorakis}, {Pellat}, {Scudder}, and
  {Sonnerup}}]{reme01:_first_clust_cis}
{R{\`e}me}, H., {Aoustin}, C., {Bosqued}, J.~M., {Dandouras}, I., {Lavraud},
  B., {Sauvaud}, J.~A., {Barthe}, A., {Bouyssou}, J., {Camus}, T.,
  {Coeur-Joly}, O., {Cros}, A., {Cuvilo}, J., {Ducay}, F., {Garbarowitz}, Y.,
  {Medale}, J.~L., {Penou}, E., {Perrier}, H., {Romefort}, D., {Rouzaud}, J.,
  {Vallat}, C., {Alcayd{\'e}}, D., {Jacquey}, C., {Mazelle}, C., {D'Uston}, C.,
  {M{\"o}bius}, E., {Kistler}, L.~M., {Crocker}, K., {Granoff}, M., {Mouikis},
  C., {Popecki}, M., {Vosbury}, M., {Klecker}, B., {Hovestadt}, D., {Kucharek},
  H., {Kuenneth}, E., {Paschmann}, G., {Scholer}, M., {Sckopke}, N.,
  {Seidenschwang}, E., {Carlson}, C.~W., {Curtis}, D.~W., {Ingraham}, C.,
  {Lin}, R.~P., {McFadden}, J.~P., {Parks}, G.~K., {Phan}, T., {Formisano}, V.,
  {Amata}, E., {Bavassano-Cattaneo}, M.~B., {Baldetti}, P., {Bruno}, R.,
  {Chionchio}, G., {di Lellis}, A., {Marcucci}, M.~F., {Pallocchia}, G.,
  {Korth}, A., {Daly}, P.~W., {Graeve}, B., {Rosenbauer}, H., {Vasyliunas}, V.,
  {McCarthy}, M., {Wilber}, M., {Eliasson}, L., {Lundin}, R., {Olsen}, S.,
  {Shelley}, E.~G., {Fuselier}, S., {Ghielmetti}, A.~G., {Lennartsson}, W.,
  {Escoubet}, C.~P., {Balsiger}, H., {Friedel}, R., {Cao}, J.-B., {Kovrazhkin},
  R.~A., {Papamastorakis}, I., {Pellat}, R., {Scudder}, J., and {Sonnerup}, B.:
  {First multispacecraft ion measurements in and near the Earth's magnetosphere
  with the identical Cluster ion spectrometry (CIS) experiment}, \anngeo, 19,
  1303--1354, 2001.

\bibitem[{{Scholer} et~al.(1993){Scholer}, {Fujimoto}, and
  {Kucharek}}]{scholer93:_two}
{Scholer}, M., {Fujimoto}, M., and {Kucharek}, H.: {Two-dimensional simulations
  of supercritical quasi-parallel shocks: upstream waves, downstream waves, and
  shock re-formation}, \jgr, 98, 18\,971, 1993.

\bibitem[{{Schwartz} and
  {Burgess}(1991)}]{schwartz91:_quasi_paral_shock_patch_of}
{Schwartz}, S.~J. and {Burgess}, D.: {Quasi-parallel shocks - A patchwork of
  three-dimensional structures}, \grl, 18, 373--376, 1991.

\bibitem[{{Schwartz} et~al.(1985){Schwartz}, {Chaloner}, {Hall},
  {Christiansen}, and {Johnstones}}]{schwartz85}
{Schwartz}, S.~J., {Chaloner}, C.~P., {Hall}, D.~S., {Christiansen}, P.~J., and
  {Johnstones}, A.~D.: {An active current sheet in the solar wind}, \nat, 318,
  269--271, 1985.

\bibitem[{{Schwartz} et~al.(2000){Schwartz}, {Paschmann}, {Sckopke}, {Bauer},
  {Dunlop}, {Fazakerley}, and {Thomsen}}]{schwartz00:_condit}
{Schwartz}, S.~J., {Paschmann}, G., {Sckopke}, N., {Bauer}, T.~M., {Dunlop},
  M., {Fazakerley}, A.~N., and {Thomsen}, M.~F.: {Conditions for the formation
  of hot flow anomalies at Earth's bow shock}, \jgr, 105, 12\,639--12\,650,
  \doi{10.1029/1999JA000320}, 2000.

\bibitem[{{Sibeck} et~al.(1991){Sibeck}, {Lopez}, and
  {Roelof}}]{sibeck91:_solar}
{Sibeck}, D.~G., {Lopez}, R.~E., and {Roelof}, E.~C.: {Solar wind control of
  the magnetopause shape, location, and motion}, \jgr, 96, 5489--5495, 1991.

\bibitem[{{Sibeck} et~al.(1999){Sibeck}, {Borodkova}, {Schwartz}, {Owen},
  {Kessel}, {Kokubun}, {Lepping}, {Lin}, {Liou}, {L{\"u}hr}, {McEntire},
  {Meng}, {Mukai}, {Nemecek}, {Parks}, {Phan}, {Romanov}, {Safrankova},
  {Sauvaud}, {Singer}, {Solovyev}, {Szabo}, {Takahashi}, {Williams}, {Yumoto},
  and {Zastenker}}]{sibeck99:_compr}
{Sibeck}, D.~G., {Borodkova}, N.~L., {Schwartz}, S.~J., {Owen}, C.~J.,
  {Kessel}, R., {Kokubun}, S., {Lepping}, R.~P., {Lin}, R., {Liou}, K.,
  {L{\"u}hr}, H., {McEntire}, R.~W., {Meng}, C.-I., {Mukai}, T., {Nemecek}, Z.,
  {Parks}, G., {Phan}, T.~D., {Romanov}, S.~A., {Safrankova}, J., {Sauvaud},
  J.-A., {Singer}, H.~J., {Solovyev}, S.~I., {Szabo}, A., {Takahashi}, K.,
  {Williams}, D.~J., {Yumoto}, K., and {Zastenker}, G.~N.: {Comprehensive study
  of the magnetospheric response to a hot flow anomaly}, \jgr, 104, 4577--4594,
  \doi{10.1029/1998JA900021}, 1999.

\bibitem[{{Sibeck} et~al.(2002){Sibeck}, {Phan}, {Lin}, {Lepping}, and
  {Szabo}}]{sibeck02:_wind}
{Sibeck}, D.~G., {Phan}, T.-D., {Lin}, R., {Lepping}, R.~P., and {Szabo}, A.:
  {Wind observations of foreshock cavities: A case study}, \jgr, 107, 4--1,
  \doi{10.1029/2001JA007539}, 2002.

\bibitem[{{Smith} et~al.(1998){Smith}, {L'Heureux}, {Ness}, {Acu{\~n}a},
  {Burlaga}, and {Scheifele}}]{smith98:_ace_magnet_field_exper}
{Smith}, C.~W., {L'Heureux}, J., {Ness}, N.~F., {Acu{\~n}a}, M.~H., {Burlaga},
  L.~F., and {Scheifele}, J.: {The ACE Magnetic Fields Experiment}, Space
  Science Reviews, 86, 613--632, \doi{10.1023/A:1005092216668}, 1998.

\bibitem[{{Tanaka} et~al.(1983){Tanaka}, {Goodrich}, {Winske}, and
  {Papadopoulos}}]{tanaka83}
{Tanaka}, M., {Goodrich}, C.~C., {Winske}, D., and {Papadopoulos}, K.: {A
  source of the backstreaming ion beams in the foreshock region}, \jgr, 88,
  3046--3054, 1983.

\bibitem[{{Thomas}(1989)}]{thomas89:_three}
{Thomas}, V.~A.: {Three-dimensional simulation of diamagnetic cavity formation
  by a finite-size plasma beam}, \jgr, 94, 13\,579--13\,583, 1989.

\bibitem[{{Thomas} and {Brecht}(1988)}]{thomas88:_evolut}
{Thomas}, V.~A. and {Brecht}, S.~H.: {Evolution of diamagnetic cavities in the
  solar wind}, \jgr, 93, 11\,341--11\,353, 1988.

\bibitem[{{Thomas} et~al.(1991){Thomas}, {Winske}, {Thomsen}, and
  {Onsager}}]{thomas91:_hybrid}
{Thomas}, V.~A., {Winske}, D., {Thomsen}, M.~F., and {Onsager}, T.~G.: {Hybrid
  simulation of the formation of a hot flow anomaly}, \jgr, 96, 11\,625, 1991.

\bibitem[{{Thomsen} et~al.(1986){Thomsen}, {Gosling}, {Fuselier}, {Bame}, and
  {Russell}}]{thomsen86:_hot}
{Thomsen}, M.~F., {Gosling}, J.~T., {Fuselier}, S.~A., {Bame}, S.~J., and
  {Russell}, C.~T.: {Hot, diamagnetic cavities upstream from the earth's bow
  shock}, \jgr, 91, 2961--2973, 1986.

\bibitem[{{Thomsen} et~al.(1993){Thomsen}, , {Thomas}, {Winske}, {Gosling},
  {Farris}, and {Russell}}]{thomsen93:_obser_test_hot_flow_abnom}
{Thomsen}, M.~F., , {Thomas}, V.~A., {Winske}, D., {Gosling}, J.~T., {Farris},
  M.~H., and {Russell}, C.~T.: Observational Test of Hot Flow Anomaly Formation
  by the Interaction of a Magnetic Discontinuity With the Bow Shock, \jgr, 98,
  15\,319--15\,330, 1993.

\bibitem[{{Tsyganenko}(1995)}]{tsyganenko95:_model}
{Tsyganenko}, N.~A.: {Modeling the Earth's magnetospheric magnetic field
  confined within a realistic magnetopause}, \jgr, 100, 5599--5612, 1995.

\end{thebibliography}

\end{document}